\documentclass[12pt, draftclsnofoot, onecolumn]{IEEEtran}
\usepackage{amsmath,amsthm,amsfonts,amscd,amssymb,graphics,dsfont}
\usepackage[mathscr]{eucal}
\usepackage{graphics,graphicx,multicol}
\usepackage{epsfig}
\usepackage{epstopdf}
\usepackage{subfigure}
\usepackage{stfloats}
\usepackage{latexsym}
\usepackage{amsfonts}
\usepackage{cite}
\usepackage{algorithm,algorithmic}
\usepackage{color,xcolor}
\usepackage{multirow}

\newtheorem{thm}{Theorem}
\newtheorem{defn}{Definition}

\begin{document}
\title{Task-Oriented Edge Networks: Decentralized Learning Over Wireless Fronthaul}
\author{Hoon Lee,~\IEEEmembership{Member, IEEE}, and Seung-Wook Kim,~\IEEEmembership{Member, IEEE}
\thanks{
H. Lee is with the Department of Electrical Engineering and the Artificial Intelligence Graduate School, Ulsan National Institute of Science and Technology (UNIST), Ulsan, 44919, South Korea (e-mail: hoonlee@unist.ac.kr).

S.-W. Kim is with the Division of Electrical and Communication Engineering, Pukyong National University, Busan 48513, South Korea (e-mail: swkim@pknu.ac.kr).

© 2023 IEEE. Personal use of this material is permitted.  Permission from IEEE must be obtained for all other uses, in any current or future media, including reprinting/republishing this material for advertising or promotional purposes, creating new collective works, for resale or redistribution to servers or lists, or reuse of any copyrighted component of this work in other works.}
} 

\maketitle
\begin{abstract}
This paper studies task-oriented edge networks where multiple edge internet-of-things nodes execute machine learning tasks with the help of powerful deep neural networks (DNNs) at a network cloud. Separate edge nodes (ENs) result in a partially observable system where they can only get partitioned features of the global network states. These local observations need to be forwarded to the cloud via resource-constrained wireless fronthual links. Individual ENs compress their local observations into uplink fronthaul messages using task-oriented encoder DNNs. Then, the cloud carries out a remote inference task by leveraging received signals. Such a distributed topology requests a decentralized training and decentralized execution (DTDE) learning framework for designing edge-cloud cooperative inference rules and their decentralized training strategies. First, we develop fronthaul-cooperative DNN architecture along with proper uplink coordination protocols suitable for wireless fronthaul interconnection. Inspired by the nomographic function, an efficient cloud inference model becomes an integration of a number of shallow DNNs. This modulized architecture brings versatile calculations that are independent of the number of ENs. Next, we present a decentralized training algorithm of separate edge-cloud DNNs over downlink wireless fronthaul channels. An appropriate downlink coordination protocol is proposed, which backpropagates gradient vectors wirelessly from the cloud to the ENs. 
Numerical results demonstrate the viability of the proposed DTDE framework for optimizing task-oriented edge networks. 
\end{abstract}

\section{Introduction}
Artificial intelligence (AI) technologies have brought a paradigm shift in realizing intelligent edge networks \cite{YShi:20,WXu:23,LK:22}. By means of powerful deep neural network (DNN) models installed at network clouds, it is viable to provide remote AI task execution services for edge internet-of-things (IoT) devices having limited computing resources \cite{SYun:22}. To achieve this goal, edge nodes (ENs) need to convey their own data samples to the cloud through fronthaul links that are subject to constraints on wireless time-frequency resources. This triggers recent studies on task-oriented edge networks that employ DNN-aided edge encoders to extract compressed features relevant for cloud inference \cite{YShi:23,DGunduz:23}. Unlike existing joint source-channel coding approaches \cite{OShea:17,NFarsad:18,JSCC:19,TTung:22} which train a pair of neural encoder-decoder to enhance the communication performance, the task-oriented network aims at maximizing the AI task execution performance at the cloud. By doing so, we can obtain task-oriented edge encoding strategies along with optimized cloud inference models. 

The optimization problems of task-oriented edge networks involve a joint design of neural edge encoders and cloud inference models that collaboratively estimate target labels through resource-constrained fronthaul interconnections. In addition, the inference and training processes of edge encoder DNNs and cloud DNN should be executed in a decentralized manner over imperfect fronthaul links. Conventional studies have focused on designing decentralized inference while assuming ideal centralized training procedures through noiseless fronthaul channels. Furthermore, the generalization ability to handle arbitrary EN populations has not yet been studied adequately. As a result, existing works require a number of edge/cloud DNNs dedicated to all possible edge network configurations. Such limitations prohibit the real-world implementation of task-oriented edge networks. To tackle these difficulties, this paper develops decentralized and versatile learning strategies for task-oriented edge networks with wireless fronthaul channels and arbitrary EN populations.

\subsection{Motivations and Related Works}

Cooperative edge-cloud DNN architectures were proposed for the task-oriented edge networks to execute remote AI inference tasks, such as network management \cite{HLee:19,HLee:21a,HLee:22,ZWang:22}, image classification \cite{JShao:22,JShao:23,SXie:23,YKim:23,XXu:23,MJan:21}, natural language processing \cite{HXie:22}, and video analysis applications \cite{JShao:23b}. 
Existing works have been confined to a \textit{centralized training and decentralized execution (CTDE)} setup, which trains a group of neural edge encoders as well as the cloud DNN model centrally. Trained DNNs are then employed at dedicated nodes for the decentralized edge-cloud inference. However, this method brings prohibitive fronthaul signaling overheads in the training phase, requiring a centralized data collection step from all ENs to the cloud. 
For this reason, a \textit{decentralized training and decentralized execution (DTDE)} framework \cite{SHwang:22} plays a significant role in designing practical task-oriented edge networks. 
Along with decentralized edge-to-cloud cooperative inference strategies, we need to develop cloud-to-edge fronthaul coordination protocols that facilitate decentralized backpropagation processes to train edge encoders and cloud DNN. This requires a joint design of uplink (edge-to-cloud inference) and downlink (cloud-to-edge backpropagation) interaction protocols to maximize desired AI task performance.

Another challenge stems from the robustness to channel impairments in the fronthaul coordination, such as resource constraints and random fading coefficients.
To accommodate capacity-constrained fronthaul links, neural edge quantization techniques were presented \cite{HLee:19,HLee:21a,JShao:22,JShao:23,JShao:23b,SXie:23} while assuming noiseless fronthaul coordination. The impact of channel imperfections has been recently incorporated in the DNN construction \cite{HLee:21a,YKim:23,HXie:22}. The additive Gaussian noise channels were taken into account \cite{HLee:21a,YKim:23}, and the works in \cite{HXie:22,ZZhang:22,XZeng:22,XXu:23,MJan:21} injected the Rayleigh fading channels into the edge-to-cloud uplink fronthaul links. These existing studies are, however, limited to the CTDE setup. The DTDE policy has been recently investigated in line with the vertical federated learning (VFL) framework \cite{YHu:19,JZhang:22,TChen:20}, but under ideal noiseless fronthaul links. It is still unaddressed how to tackle the channel impairments in the downlink fronthaul coordination to perform valid gradient calculations over noisy edge-cloud interactions.

Future edge network trends toward massive connectivity services of a number of edge IoT devices. Such a feature requests scalable DTDE architectures for the task-oriented edge networks whose inference and training calculations become independent of the number of ENs. Since typical DNNs work only with fixed input and output dimensions, they lack the versatile computation ability for arbitrary given EN populations. For this reason, single EN systems in \cite{JShao:22,SXie:23,XXu:23,MJan:21} cannot be straightforwardly applied to multi-EN task-oriented networks. The cloud DNN models provided in \cite{TChen:20,JZhang:22,JShao:23} accept a concatenated vector of encoded signals sent by all ENs. Thus, a cloud DNN trained at a certain EN population cannot be directly applied to other network configurations with different numbers of ENs. This issue can be tackled by the sum-pooling operation \cite{YHu:19,XZeng:22,YKim:23} where the aggregation of all received signals is exploited for the inference at the cloud. However, this approach cannot control the compression rate at the ENs as the output dimension of encoder DNNs should be equal to that of the target label. This motivates us to build an appropriate cloud DNN architecture as well as its learning policy that is suitable for handling multiple edge-encoded signals sent by arbitrary EN populations.

\subsection{Contributions}
This paper proposes a DTDE learning framework for task-oriented edge networks where a cloud interacts with a group of ENs through wireless fronthaul links to execute its inference model. Separate ENs can only get access to their local data observations, which are regarded as partitioned features of the global network state. For the communication-efficient coordination under resource-constrained fronthaul links, each EN leverages an encoder DNN which generates a compressed fronthaul message. The resulting encoded signals are then sent to the cloud through uplink fronthaul channels corrupted by multiplicative fading and additive noise. With these partitioned and noisy received signals at hand, the cloud infers desired outputs using its DNN model. Accurate inference of the cloud DNN resorts to full knowledge of all partitioned information sent by the ENs. Thus, a naive approach is to utilize a concatenation of received signals as an input feature \cite{TChen:20,JZhang:22,JShao:23}. This, however, leads to a rigid structure that works only for a certain number of ENs.

To build an efficient cloud DNN model, we first interpret an oracle edge-cloud inference rule as the nomographic function. The oracle inference model can be decomposed into a set of decentralized edge encoding functions followed by a sum-pooling layer together with a cloud inference function. This results in sum-aggregation-based cloud DNN models that have been widely adopted in the VFL \cite{YHu:19,JZhang:22,CXie:22} and task-oriented networks \cite{JShao:23b,YShi:23b}. However, such an interpretation is shown to be no longer valid for wireless fronthaul channels that incur random amplitude changes to the output of the edge encoder DNNs, possibly losing the optimality. This issue can be resolved by leveraging the Kolmogorov-Arnold (KA) representation \cite{KA} which allows arbitrary continuous-valued edge encoding and wireless fronthaul channels. As a result, the cloud DNN is built as an integration of several shallow DNN modules, offering a scalable architecture whose computations are irrelevant to the number of ENs. 
By doing so, we can establish fronthaul-cooperative decentralized inference rules viable for practical task-oriented edge networks.

Next, we present decentralized training policies where the cloud and ENs collaboratively optimize their DNNs over wireless fronthaul channels. The joint design of uplink-downlink fronthaul coordination plays a critical role in the proposed training mechanism. Our careful investigations reveal that the stochastic gradient descent (SGD) algorithm for the edge encoder DNNs can be obtained at individual ENs with the help of message propagation mechanisms at the cloud. Downlink fronthaul coordination protocols are proposed where the cloud transfers downlink messages encapsulating the gradient information to the ENs over wireless fronthaul channels. These message vectors are then exploited at each EN to perform local SGD updates of its encoder DNN. Consequently, we can decouple the SGD updates of the cloud DNN and individual edge encoder DNNs, thereby leading to the decentralized training mechanism.

Local observations of the ENs would share an identical knowledge basis or have a similar modality. In this case, the effectiveness of the edge encoder DNNs can be further improved by leveraging the parameter sharing technique, which forces ENs to reuse the identical encoder DNN. As a result, the shared encoder DNN can infer the global input data by observing the partitioned information vectors of all ENs. Furthermore, such an approach leads to a scalable edge encoder architecture where a sole encoder DNN is reused across the entire ENs. As a consequence, the proposed learning structure can be applied to arbitrary task-oriented edge networks with a random EN population. 
The performance of the proposed task-oriented edge networking strategies is examined over classification tasks of various image datasets such as Tiny ImageNet~\cite{TinyImageNet} and Food-101~\cite{FOOD101}. Numerical results validate the effectiveness of the proposed approach over conventional methods.

The contributions of this work are summarized as follows:
\begin{itemize}
    \item We design a versatile and decentralized inference structure for task-oriented edge networks in the presence of wireless fronthaul channels. To this end, we exploit the KA representation theorem, which decomposes an oracle cloud inference into a group of component DNNs at the cloud along with decentralized edge encoder DNNs. Such a modulized architecture leads to the scalable cloud DNN model whose computations are independent of the number of ENs.
    \item The decentralized training strategy over wireless fronthaul channels is presented where the cloud and ENs collaboratively optimize their DNNs only with limited information sharing. We develop joint uplink-downlink communication protocols for realizing the proposed decentralized training algorithm. As a result, gradient vectors required at individual ENs can be propagated successfully through downlink wireless fronthaul channels.
    \item To further enhance the scalability, the encoder sharing mechanism is provided, which forces all ENs to utilize the identical encoder DNN. This invokes a new challenge in developing decentralized training policies satisfying the consensus constraints on edge encoder DNNs. We address this issue by incorporating additional gradient aggregation steps in the cloud. By doing so, the shared encoder DNN can be optimized in a decentralized manner.
    \item The viability of the proposed framework is demonstrated for various image classification tasks such as Tiny ImageNet and Food-101 datasets. Intensive simulation results demonstrate the superiority of the proposed approach to existing models in terms of the accuracy performance and scalability.
\end{itemize}

\subsection{Organization and Notations}
This paper is organized as follows. Section \ref{sec:sec2} describes the system model for task-oriented edge networks. Section \ref{sec:sec3} proposes the cloud DNN architecture and the fronthaul-cooperative decentralized inference policy. In Section \ref{sec:sec4}, we present uplink-downlink fronthaul interaction protocols that facilitate decentralized training of edge-cloud DNNs. Section \ref{sec:sec5} introduces several extension approaches to practical scenarios. Numerical results validating the proposed framework are presented in Section \ref{sec:sec6}. Finally, Section \ref{sec:sec7} concludes the paper.

\textit{Notations:} Uppercase boldface, lowercase boldface, and normal letters denote matrices, column vectors, and scalars, respectively. Sets of $p$-by-$q$ complex- and real-valued matrices are expressed as $\mathbb{C}^{p\times q}$ and $\mathbb{R}^{p\times q}$, respectively, whereas $\mathbb{C}^{p}$ and $\mathbb{R}^{p}$ respectively stand for sets of complex- and real-valued column vectors of length $p$. The element-wise multiplication operator of vectors is defined as $\odot$. A set of integers $\{a,a+1,\cdots,b-1,b\}$ is denoted by $[a,b]$. The all-zero column vector of length $p$ is expressed as $\mathbf{0}_{p}$, and $\mathbf{I}_{p}$ denotes the identity matrix of size $p$-by-$p$. Also, $\nabla_{V}$ accounts for the gradient operator with respect to a variable $V$.

\section{System Model} \label{sec:sec2}

\begin{figure}
\centering
\includegraphics[width=.7\linewidth]{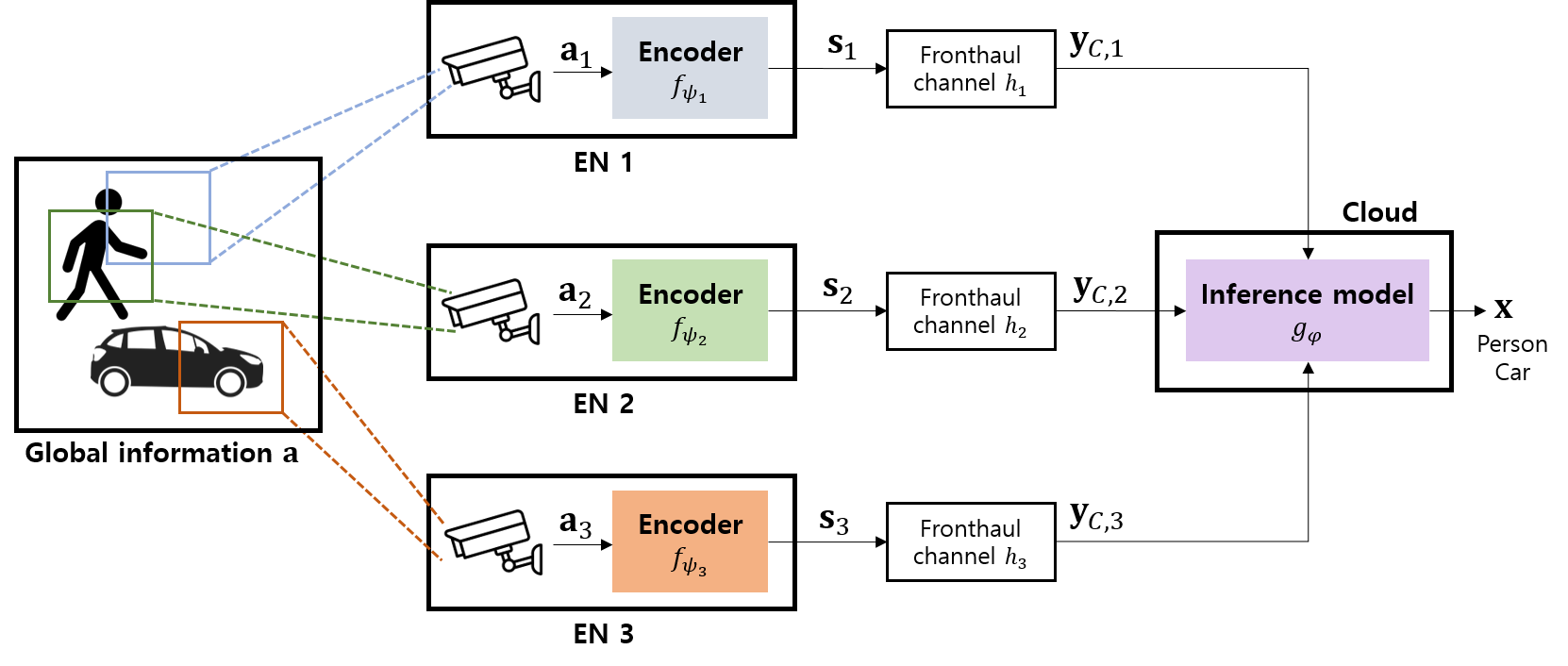}
\caption{Task-oriented edge network with $N=3$ ENs.}
\label{fig:fig1}
\end{figure}

We consider a task-oriented edge network in Fig. \ref{fig:fig1} where a cloud carries out remote AI task computations of $N$ ENs connected via dedicated fronthaul links. Let $\mathcal{N}\triangleq[1,N]$ be the set of indices of ENs. The ENs aim at executing AI computation tasks for global information $\mathbf{a}$ given to the overall edge network. Since the ENs are separated over the network area, each EN $i$ can only get its own local observation $\mathbf{a}_{i}\in\mathbb{R}^{A}$ of length $A$ using embedded sensors, e.g., cameras. For this reason, these local observations are regarded as partitions of the global network information $\mathbf{a}$. For instance, $\mathbf{a}$ would be a full-size image of the target area, and the local observations become randomly cropped images of $\mathbf{a}$ possibly overlapping each other.
By collecting these local observations, the cloud estimates an output $\mathbf{x}\in\mathbb{R}^{X}$ of length $X$ by using its DNN $g_{\varphi}$ with trainable parameter $\varphi$. 

However, conveying a raw measurement $\mathbf{a}_{i}$ directly to the cloud is not viable due to limited fronthaul resources. This invokes decentralized edge encoding processes to create compressed message vectors. Without loss of the generality, each fronthaul link is assumed to convey $\tilde{S}\triangleq S/2$ ($S\ll A$) complex numbers per channel use, e.g., $\tilde{S}$ time-frequency resource blocks (RBs) are assigned to each fronthaul link. Each EN $i$ encodes its local observation $\mathbf{a}_{i}$ into a real-valued fronthaul message $\mathbf{s}_{i}\in\mathbb{R}^{S}$ of length $S$, where each element of $\mathbf{s}_{i}$ occupies one fronthaul RB. Let $f_{\psi_{i}}:\mathbb{R}^{A}\rightarrow\mathbb{R}^{S}$ be an encoder DNN of EN $i$ where $\psi_{i}$ accounts for a trainable parameter. Thus, EN $i$ creates its message vector $\mathbf{s}_{i}\in\mathbb{R}^{S}$~as
\begin{align}
    \mathbf{s}_{i}=f_{\psi_{i}}(\mathbf{a}_{i}).\label{eq:si}
\end{align}
A group of edge encoder DNNs $f_{\psi_{i}}$, $\forall i\in\mathcal{N}$, are optimized together with the cloud DNN $g_{\varphi}$ in a task-oriented manner such that the cloud inference performance is maximized.

For the baseband signal processing, we represent the real-valued message vector $\mathbf{s}_{i}$ as
\begin{align}
    \mathbf{s}_{i}=[\mathbf{s}_{R,i}^{T},\mathbf{s}_{I,i}^{T}]^{T},\label{eq:sibb}
\end{align}
where $\mathbf{s}_{R,i}\in\mathbb{R}^{\tilde{S}}$ and $\mathbf{s}_{I,i}\in\mathbb{R}^{\tilde{S}}$ respectively stand for real and imaginary parts of a complex-valued baseband signal vector $\tilde{\mathbf{s}}_{i}\in\mathbb{C}^{\tilde{S}}$ obtained as
\begin{align}
    \tilde{\mathbf{s}}_{i}=\mathbf{s}_{R,i}+j\mathbf{s}_{I,i}.
\end{align}
We impose the peak transmit power budget $p_{E}$ for each fronthaul RB. Let $\mathbf{v}[j]$ be the $j$-th element of a vector $\mathbf{v}$. Then, the transmit power constraint at EN $i$ is expressed by
\begin{align}\label{eq:pc}
    |\tilde{\mathbf{s}}_{i}[j]|^2 \leq p_{E}.
\end{align}
To satisfy the transmit power constraint in \eqref{eq:pc}, the output activation function of $f_{\psi_{i}}$ is set to the projection operation. Let $\mathbf{v}_{i}=[\mathbf{v}_{R,i}^{T},\mathbf{v}_{I,i}^{T}]^{T}\in\mathbb{R}^{S}$ with $\mathbf{v}_{R,i},\mathbf{v}_{I,i}\in\mathbb{R}^{\tilde{S}}$ be the output of $f_{\psi_{i}}$ before the activation. Defining $p[j]\triangleq|\mathbf{v}_{R,i}[j]|^{2}+|\mathbf{v}_{I,i}[j]|^{2}$, 
The output activation function of $f_{\psi_{i}}$ yields the final output $\mathbf{s}_{i}=[\mathbf{s}_{R,i}^{T},\mathbf{s}_{I,i}^{T}]^{T}$ as
\begin{align}\label{eq:proj}
    \mathbf{s}_{\sf{X},i}[j]=
    \begin{cases}
        \mathbf{v}_{\sf{X},i}[j] & \text{if }p[j] \leq p_{E}\\
        \sqrt{\frac{p_{E}}{p[j]}}\mathbf{v}_{\sf{X},i}[j] & \text{elsewhere}
    \end{cases}
\end{align}
for $\mathsf{X}\in\{R,I\}$.

The edge-encoded messages $\tilde{\mathbf{s}}_{i}$, $\forall i\in\mathcal{N}$, are sent to the cloud through orthogonal fronthaul RBs. The uplink fronthaul interaction from EN $i$ to the cloud is corrupted by the additive Gaussian noise $\tilde{\mathbf{n}}_{C,i}\sim\mathcal{CN}(\mathbf{0}_{\tilde{S}},\sigma_{C}^{2}\mathbf{I}_{\tilde{S}})\in\mathbb{C}^{\tilde{S}}$ with variance $\sigma_{C}^{2}$ and the fading $\tilde{\mathbf{h}}_{i}\sim\mathcal{CN}(\mathbf{0}_{\tilde{S}},\mathbf{I}_{\tilde{S}})$ given~as
\begin{align}
    \tilde{\mathbf{h}}_{i}=|\tilde{\mathbf{h}}_{i}|\odot e^{\angle\tilde{\mathbf{h}}_{i}}, \label{eq:htilde}
\end{align}
where $|\cdot|$ and $\angle\cdot$ respectively indicate element-wise absolute and angle operators, $|\tilde{\mathbf{h}}_{i}|\in\mathbb{R}^{\tilde{S}}$ and $\angle\tilde{\mathbf{h}}_{i}\in\mathbb{R}^{\tilde{S}}$ indicate the amplitude and phase of the channel $\tilde{\mathbf{h}}_{i}$, respectively.

EN $i$ employs a linear precoding strategy to mitigate the phase ambiguity. The precoded signal of EN $i$ is written~by 
\begin{align}
    e^{-\angle\tilde{\mathbf{h}}_{i}}\odot\tilde{\mathbf{s}}_{i}. \label{eq:sip}
\end{align}
The received signal vector at the cloud, denoted by $\tilde{\mathbf{y}}_{C,i}\in\mathbb{C}^{\tilde{S}}$, is obtained as
\begin{align}\label{eq:ytilde}
    \tilde{\mathbf{y}}_{C,i}
    =\tilde{\mathbf{h}}_{i}\odot e^{-\angle\tilde{\mathbf{h}}_{i}}\odot\tilde{\mathbf{s}}_{i}+\tilde{\mathbf{n}}_{C,i}=|\tilde{\mathbf{h}}_{i}|\odot\tilde{\mathbf{s}}_{i}+\tilde{\mathbf{n}}_{C,i},
\end{align}
where elements of the effective channel gain vector $|\tilde{\mathbf{h}}_{i}|$ follow the Rayleigh distribution.
For convenience, we represent the complex-valued signal model in \eqref{eq:ytilde} with real-valued vectors $\mathbf{y}_{C,i}\in\mathbb{R}^{S}$ and $\mathbf{n}_{C,i}\in\mathbb{R}^{S}$ defined as
\begin{align}
    \mathbf{y}_{C,i}=[\Re\{\tilde{\mathbf{y}}_{C,i}^{T}\},\Im\{\tilde{\mathbf{y}}_{C,i}^{T}\}]^{T},\ \mathbf{n}_{C,i}=[\Re\{\tilde{\mathbf{n}}_{C,i}^{T}\},\Im\{\tilde{\mathbf{n}}_{C,i}^{T}\}]^{T},
\end{align}
where $\Re\{v\}$ and $\Im\{v\}$ equal real and imaginary parts of a complex number $v$, respectively. Then, the equivalent real representation of \eqref{eq:ytilde} becomes
\begin{align}
    \mathbf{y}_{C,i}=\mathbf{H}_{i}\mathbf{s}_{i}+\mathbf{n}_{C,i}\triangleq h_{i}(\mathbf{s}_{i}), \label{eq:yi_fading}
\end{align}
where $h_{i}:\mathbb{R}^{S}\rightarrow\mathbb{R}^{S}$ accounts for the effective channel transfer function of the fronthaul link from EN $i$ to the cloud, $\mathbf{H}_{i}\triangleq\text{diag}([|\tilde{\mathbf{h}_{i}}|^{T},|\tilde{\mathbf{h}_{i}}|^{T}]^{T})\in\mathbb{R}^{S\times S}$ is the real-valued effective channel matrix, and $\text{diag}(\mathbf{v})$ is a diagonal matrix whose diagonal entries equal to elements of a vector~$\mathbf{v}$.

Upon receiving $\mathbf{y}_{C,i}$ in \eqref{eq:yi_fading}, $\forall i\in\mathcal{N}$, the cloud infers an estimate $\mathbf{x}$ of the desired label vector $\mathbf{t}\in\mathbb{R}^{X}$ of the global input $\mathbf{a}$ using the cloud DNN $g_{\varphi}:\mathbb{R}^{NS}\rightarrow\mathbb{R}^{X}$ as
\begin{subequations}\label{eq:gy}
\begin{align}
    \mathbf{x}&=g_{\varphi}(\mathbf{y}_{C,1},\cdots,\mathbf{y}_{C,N})\\
    &=g_{\varphi}(h_{1}\circ f_{\psi_{1}}(\mathbf{a}_{1}),\cdots,h_{N}\circ f_{\psi_{N}}(\mathbf{a}_{N})), 
\end{align}
\end{subequations}
where $p\circ q$ denotes the composition of functions $p$ and $q$. The performance of the cloud DNN $g_{\varphi}$ can be measured by a loss function $l(\mathbf{x},\mathbf{t})$.
The corresponding training problem is formulated as
\begin{align}
    &\min_{\theta} L(\theta)\triangleq\mathbb{E}[l(\mathbf{x},\mathbf{t})] \label{eq:L},
\end{align}
where 
$\theta\triangleq\varphi\bigcup\{\psi_{i}:\forall i\in\mathcal{N}\}$ is the collection of all trainable parameter sets and 
the average loss function $L(\theta)$ in \eqref{eq:L} evaluates the expected performance over the joint probability distribution of the fronthaul channels $h_{i}$, $\forall i\in\mathcal{N}$, all local observations $\{\mathbf{a}_{i}:\forall i\in\mathcal{N}\}$, and the label $\mathbf{t}$.

The task-oriented edge network training formalism \eqref{eq:L} entails two-fold design challenges. Due to randomness in fronthaul channels, practical edge networks consist of an arbitrary number of active ENs. Thus, the EN population $N$ is no longer a fixed number but is given by a random variable that changes at each inference step. However, the input dimension to the cloud DNN in \eqref{eq:gy} scales with $N$, implying that a simple multi-layer perceptron (MLP) model in \cite{TChen:20,JZhang:22,JShao:23}, which accepts the concatenated vector $[\mathbf{y}_{C,1}^{T},\cdots,\mathbf{y}_{C,N}^{T}]^{T}\in\mathbb{R}^{NS}$ as an input to the cloud DNN $g_{\varphi}$, fails to establish a versatile computation structure for arbitrary $N$. Therefore, it is essential to build a proper cloud DNN that is scalable to the EN population. In addition, the distributed nature of the ENs and cloud requests a valid fronthaul interaction protocol to proceed with both the inference and training calculations. To address these challenges, we propose a novel DTDE learning framework for task-oriented edge networks.

\section{Decentralized Inference Strategy} \label{sec:sec3}

This section presents an efficient fronthaul cooperation policy for the task-oriented edge network. To this end, we first design a versatile computation structure of the cloud DNN $g_{\varphi}$ \eqref{eq:gy} that is adaptive to arbitrary given EN population $N$. It is then followed by the description of decentralized edge-cloud inference protocols.

\subsection{Scalable Architecture}
We exploit the notion of the nomographic function to identify a valid edge-cloud learning architecture. For simplicity, our discussion focuses on scalar label $t\in\mathbb{R}$ and scalar local information $a_{i}\in\mathbb{A}\subset\mathbb{R}$, $\forall i\in\mathcal{N}$, where the domain $\mathbb{A}$ of $a_{i}$ is assumed to be a compact set. The goal of the cloud DNN $g_{\varphi}$ is to learn an oracle mapping $c:\mathbb{A}^{N}\rightarrow\mathbb{R}$ that generates a label $t$ using a set of local measurements $a_{i}$, $\forall i\in\mathcal{N}$, as
\begin{align}
    t=c(a_{1},\cdots,a_{N}).\label{eq:xstar}
\end{align}
In what follows, we introduce the definition of the nomographic function, which provides a key insight into characterizing the oracle inference function $c$ using DNNs.

\begin{defn}
A function $c:\mathbb{A}^{N}\rightarrow\mathbb{R}$ is nomographic if it can be represented as
\begin{align}
    c(a_{1},\cdots,a_{N})=u\left(\sum_{i\in\mathcal{N}}v_{i}(a_{i})\right), \label{eq:nom}
\end{align}
for some mappings $u:\mathbb{R}\rightarrow\mathbb{R}$ and $v_{i}:\mathbb{R}\rightarrow\mathbb{R}$.
\end{defn}

Any nomographic function can be decomposed into outer mapping $u$ and $N$ inner mappings $v_{i}$, $\forall i\in\mathcal{N}$. It has been reported in \cite{RBuck:79} that every function can be classified as a nomographic function. This indicates that the oracle mapping $c$ in \eqref{eq:xstar} can also be regarded as the nomographic function. Therefore, the remaining work for obtaining the oracle function $c$ is to identify proper outer and inner mappings. To this end, we facilitate learnable models $u_{\lambda}$ and $v_{\mu_{i}}$ each constructed with parameters $\lambda$ and $\mu_{i}$, respectively. Then, the oracle mapping $c$ can be written as
\begin{align}
    c(a_{1},\cdots,a_{N})= u_{\lambda}\left(\sum_{i\in\mathcal{N}}v_{\mu_{i}}(a_{i})\right). \label{eq:capp}
\end{align}
Based on \eqref{eq:capp}, one can build the cloud DNN $g_{\varphi}$ as
\begin{align}
    g_{\varphi}(\mathbf{y}_{C,1},\cdots,\mathbf{y}_{C,N})=u_{\lambda}\left(\sum_{i\in\mathcal{N}}h_{i}\circ f_{\psi_{i}}(\mathbf{a}_{i})\right), \label{eq:vim_ori}
\end{align}
where the inner mapping $v_{\mu_{i}}$ is interpreted as the output of the fronthaul channel $h_{i}$, i.e., $v_{\mu_{i}}=h_{i}\circ f_{\psi_{i}}$ with $\mu_{i}=\psi_{i}$, and the outer mapping $u_{\lambda}$ can be viewed as the cloud DNN. 

The above structure reveals that it suffices for the cloud DNN to take the aggregated received signal $\sum_{i\in\mathcal{N}}\mathbf{y}_{C,i}$ instead of processing the concatenation $[\mathbf{y}_{C,1}^{T},\cdots,\mathbf{y}_{C,N}^{T}]^{T}$. As a result, the input dimension of $g_{\varphi}$ becomes irrelevant to the EN population $N$, thereby leading to the versatile structure. As will be discussed in Sec. \ref{sec:related}, \eqref{eq:vim_ori} has been widely adopted in existing works \cite{YHu:19,JZhang:22,CXie:22,JShao:23b,YShi:23b}. The factorization in \eqref{eq:nom} is, in general, invalid for continuous mappings $u$ and $v_{i}$ \cite{RCBuck:82}. Such a restriction requires digital encoder DNNs \cite{HLee:19,HLee:21a,JShao:22,JShao:23,SXie:23} that learn discrete-valued encoding functions $f_{\psi_{i}}$, $\forall i\in\mathcal{M}$. However, when it comes to the wireless fronthaul channels $h_{i}$, the resulting inner mapping $v_{\mu_{i}}=h_{i}\circ f_{\psi_{i}}$ is no longer the discrete function due to arbitrary fading $\mathbf{H}_{i}$ and noise $\mathbf{n}_{C,i}$. 
To address these difficulties, we exploit the following KA representation theorem \cite{KA}, which removes the dependency on the discrete-valued mappings.

\begin{thm}
Let $\mathcal{M}\triangleq[1,M]$ for some integer $M$. A continuous function $c:\mathbb{A}^{N}\rightarrow\mathbb{R}$ can be represented as
\begin{align}
c(a_{1},\cdots,a_{N})=\sum_{m\in\mathcal{M}}u_{m}\left(\sum_{i\in\mathcal{N}}v_{mi}(a_{i})\right) \label{eq:KA}
\end{align}
for some continuous functions $u_{m}$, $\forall m\in\mathcal{M}$, and $v_{mi}$, $\forall (m,i)\in\mathcal{M}\times\mathcal{N}$. The outer mapping $u_{m}$ depends on $c$, whereas the choice of the inner mapping $v_{mi}$ becomes independent of $c$.
\end{thm}

This theorem states that every continuous function can be expressed as the superposition of $M$ nomopraphic functions $u_{m}(\sum_{i\in\mathcal{N}}v_{mi}(a_{i}))$, $\forall m\in\mathcal{M}$, each comprising an outer mapping $u_{m}$ and inner mappings $v_{mi}$, $\forall i\in\mathcal{N}$. These mappings can have arbitrary structures with continuous-valued inputs and outputs. As shown in \eqref{eq:vim_ori}, the inner mapping $v_{mi}$ is closely related to the fronthaul channel $h_{i}$ and encoder DNN $f_{\psi_{i}}$. The existence of the continuous inner mapping indicates that \eqref{eq:KA} suits edge-cloud collaboration over arbitrary wireless fronthaul channels. Also, we can adopt the analog transmission strategy \cite{YKim:23,HXie:22} where the encoder DNN $f_{\psi_{i}}$ determines the continuous-valued message $\mathbf{s}_{i}$ straightforwardly. This approach becomes more suitable for standard gradient-based training algorithms compared to the quantization process with invalid gradient computations \cite{HLee:19,HLee:21a,JShao:23,SXie:23}. 
Another important property of the inner mapping $v_{mi}$ is that its design procedure is independent of the oracle inference $c$. This implies that the random fronthaul channel $h_{i}$ has no critical impact on the optimization of $v_{mi}$, in particular, the encoder DNN $f_{\psi}$. 

Likewise \eqref{eq:capp}, a tractable approach for identifying $u_{m}$ and $v_{mi}$ is to employ trainable functions $u_{\lambda_{m}}$ and $v_{\mu_{mi}}$ with $\lambda_{m}$ and $\mu_{mi}$ being the trainable parameters. Hence, the cloud DNN $g_{\varphi}$ in \eqref{eq:vim_ori} can be modified as
\begin{align}
    g_{\varphi}(\mathbf{y}_{C,1},\cdots,\mathbf{y}_{C,N})
    &=\sum_{m\in\mathcal{M}}u_{\lambda_{m}}\left(\sum_{i\in\mathcal{N}}z_{\zeta_{m}}\circ h_{i}\circ f_{\psi_{i}}(\mathbf{a}_{i})\right) \label{eq:KA2},
\end{align}
where a group of component DNNs $u_{\lambda_{m}}$ and $z_{\zeta_{m}}$, $\forall m\in\mathcal{M}$, is combined into the cloud DNN $g_{\varphi}$ whose trainable parameter $\psi$ becomes $\varphi=\{(\lambda_{m},\zeta_{m}):\forall m\in\mathcal{M}\}$. In \eqref{eq:KA2}, we introduce additional trainable functions $z_{\zeta_{m}}$, $\forall m\in\mathcal{M}$, each having parameter set $\zeta_{m}$, to construct the inner mapping $v_{\mu_{mi}}$ as
\begin{align}
    v_{\mu_{mi}}\triangleq z_{\zeta_{m}}\circ h_{i} \circ f_{\psi_{i}}\label{eq:vmi}
\end{align}
with $\mu_{mi}\triangleq(\zeta_{m},\psi_{i})$ being the set of trainable parameters $\zeta_{m}$ and $\psi_{i}$. Such a design approach allows to express $MN$ independent inner mappings $v_{\mu_{mi}}$, $\forall (i,m)\in\mathcal{N}\times\mathcal{M}$, by only using $M$ DNN modules $z_{\zeta_{m}}$, $\forall m\in\mathcal{M}$. As a result, the proposed cloud DNN consists of $M$ component DNNs $u_{\lambda_{m}}$ and $z_{\zeta_{m}}$, $\forall m\in\mathcal{M}$, which is not dependent on the number of the ENs $N$. By doing so, we can establish the scalable inference model that can be applied to arbitrary given $N$.

The proposed cloud DNN architecture in \eqref{eq:KA2} successfully achieves the scalable property with respect to the number of ENs. However, the dedicated edge encoder DNNs $f_{\psi_{i}}$ still require rigid signal processing architectures at the ENs whose computations, in particular, the training phase, are dependent on the EN population. This challenge can be resolved by allowing the ENs to reuse the identical encoder DNNs. Such an encoder sharing policy will be presented in Section \ref{sec:encoder_sharing}.

\subsection{Cooperative Inference}
\begin{figure}
\centering
\includegraphics[width=.7\linewidth]{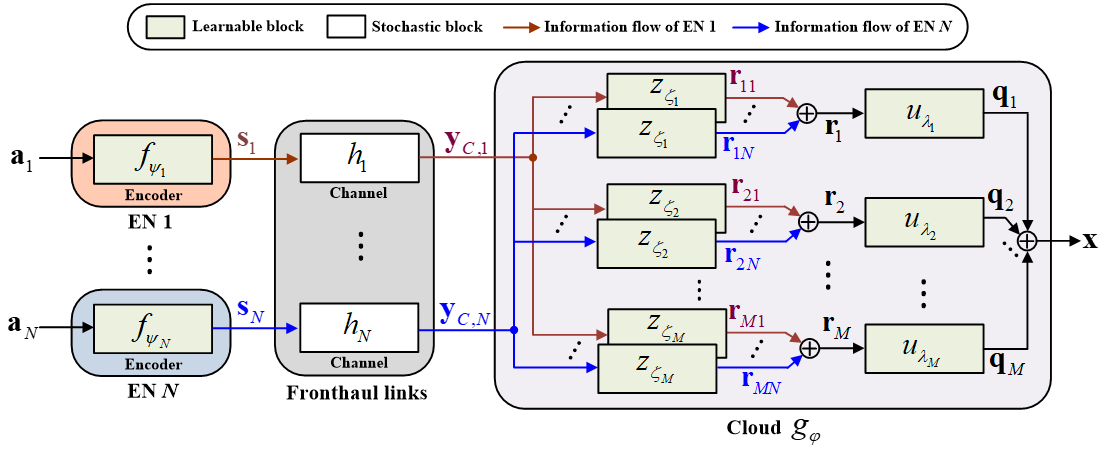}
\caption{Decentralized inference of proposed fronthaul cooperation strategy.}
\label{fig:fig2}
\end{figure}

The proposed cooperative inference architecture is illustrated in Fig. \ref{fig:fig2}, which involves decentralized edge encoder DNNs $f_{\psi_{i}}$, data transmission steps over wireless fronthaul channels $h_{i}$, and the final inference at the cloud with the DNN $g_{\varphi}$. The cloud first handles the received signals $\mathbf{y}_{C,i}$, $\forall i\in\mathcal{N}$, using $M$ component DNNs 
$z_{\zeta_{m}}:\mathbb{R}^{S}\rightarrow\mathbb{R}^{R}$, $\forall m\in\mathcal{M}$, as
\begin{align}
\mathbf{r}_{mi}=z_{\zeta_{m}}(\mathbf{y}_{i})=z_{\zeta_{m}}\circ h_{i} \circ f_{\psi_{i}}(\mathbf{a}_{i}), \label{eq:rmi}
\end{align}
where $\mathbf{r}_{mi}\in\mathbb{R}^{R}$ of length $R$ stands for the output of the $m$-th branch module $z_{\zeta_{m}}$. 
Then, the cloud aggregates intermediate outputs $\mathbf{r}_{mi}$, $\forall i\in\mathcal{N}$, using the sum-pooling operation as
\begin{align}
\mathbf{r}_{m}=\sum_{i\in\mathcal{N}}\mathbf{r}_{mi}=\sum_{i\in\mathcal{N}}z_{\zeta_{m}}(\mathbf{y}_{C,i}), \label{eq:rm}
\end{align}
where $\mathbf{r}_{m}\in\mathbb{R}^{R}$ indicates the output of the sum-pooling layer of the $m$-th branch module $z_{\zeta_{m}}$. Each $\mathbf{r}_{m}$ is further processed by another component DNN $u_{\lambda_{m}}:\mathbb{R}^{R}\rightarrow\mathbb{R}^{X}$ as
\begin{align}
\mathbf{q}_{m}=u_{\lambda_{m}}(\mathbf{r}_{m})=u_{\lambda_{m}}\left(\sum_{i\in\mathcal{N}}\mathbf{r}_{mi}\right),\label{eq:qm}
\end{align}
where $\mathbf{q}_{m}\in\mathbb{R}^{X}$ accounts for the post-processed information vector. Finally, the cloud applies the sum-pooling operation to create the estimate $\mathbf{x}$ as
\begin{align}
\mathbf{x}=\sum_{m\in\mathcal{M}}\mathbf{q}_{m}=\sum_{m\in\mathcal{M}}u_{\lambda_{m}}(\mathbf{r}_{m}).\label{eq:x3}
\end{align}

The multi-branch architecture of the proposed cloud DNN can be viewed as an extension of famous structures such as the multi-head attention layer of the transformer \cite{TRANSFORMER} and the basic block of the ResNeXt \cite{RESNEXT}. These models apply a number of independent neural transformations, e.g., the component DNNs $z_{\zeta_{m}}$, $\forall m\in\mathcal{M}$, to the low-dimensional input vectors, and then aggregates resulting outputs by using the sum-pooling operation. These architectures were developed to process a single input vector only, whereas the proposed cloud DNN handles multiple received signals $\mathbf{y}_{C,i}$, $\forall i\in\mathcal{N}$. To this end, the multi-branch transformation is applied to each $\mathbf{y}_{C,i}$, and the resulting outputs are combined across the ENs as in \eqref{eq:rm}. In addition, $M$ different outputs $\mathbf{r}_{m}$, $\forall m\in\mathcal{M}$, are further processed by the multi-branch architecture comprising the outer mappings $u_{\lambda_{m}}$, $\forall m\in\mathcal{M}$, which is followed by the sum aggregation.

Such a multi-branch architecture offers multiple forward propagation paths, also known as the \textit{cardinality} of a DNN \cite{RESNEXT, RESNEST}. The number of component DNNs $M$ acts as the cardinality of the proposed cloud DNN. Increasing the cardinality can enhance the effectiveness of a DNN without increasing its overall depth and width. Employing sufficient populations of shallow component DNNs $z_{\zeta_{m}}$, $\forall m\in\mathcal{M}$, has been shown to enhance the inference performance compared to the architecture with a single cardinality, i.e., the cloud DNN accepting the concatenated received signal. A group of component DNNs $\{z_{\zeta_{m}}:\forall m\in\mathcal{M}\}$ provide diverse representations for each received signal vector $\mathbf{y}_{C,i}$, and thus we can present the overfitting issue by encouraging the cloud DNN not to focus a particular latent feature. This regularization effect enhances generalization ability and reduces sensitivity to noisy data, e.g., the fronthaul channel noise, thereby lessening the risk of over-parameterization.  

\begin{algorithm}
\caption{Proposed Cooperative Inference Strategy}
\begin{algorithmic}[1]\label{alg:alg1}
    \FOR{\textbf{each} EN $i\in\mathcal{N}$ \textbf{in parallel}}
        \STATE{EN $i$ creates $\mathbf{s}_{i}$ from \eqref{eq:si} and sends it to the cloud.}
        \STATE{Cloud receives $\mathbf{y}_{C,i}$ through fronthaul channel $h_{i}$.} 
    \ENDFOR     
    \FOR{\textbf{each} $(i,m)\in\mathcal{N}\times\mathcal{M}$ \textbf{in parallel}}
        \STATE{Cloud recovers $\mathbf{r}_{mi}$ from \eqref{eq:rmi}.}
    \ENDFOR
    \FOR{\textbf{each} $m\in\mathcal{M}$ \textbf{in parallel}}
        \STATE{Cloud aggregates $\mathbf{r}_{m}$ from \eqref{eq:rm}.}
        \STATE{Cloud calculates $\mathbf{q}_{m}$ from \eqref{eq:qm}.}
    \ENDFOR
    \STATE{Cloud obtains $\mathbf{x}$ from \eqref{eq:x3}.}
\end{algorithmic}
\end{algorithm}

Algorithm \ref{alg:alg1} summarizes the proposed decentralized edge-cloud inference. Each EN individually obtains the fronthaul message $\mathbf{s}_{i}$ using the encoder DNN $f_{\psi_{i}}$ in \eqref{eq:si}. Upon receiving $\mathbf{y}_{C,i}$, the cloud employs the multi-branch component DNNs $z_{\zeta_{m}}$ and $u_{\lambda_{m}}$, $\forall m\in\mathcal{M}$, to get the estiamte $\mathbf{x}$. These procedures can be accelerated via parallel forward-pass of component DNNs. Each step of Algorithm \ref{alg:alg1} can be realized in a decentralized manner without collecting the information vectors $\mathbf{a}_{i}$, $\forall i\in\mathcal{N}$, centrally. Hence, the estimate $\mathbf{x}$ can be attained by the cloud by means of the decentralized uplink fronthaul coordination.  

\section{Decentralized Training Strategy} \label{sec:sec4}

This section presents a joint training algorithm of the encoder DNNs $f_{\psi_{i}}$, $\forall i\in\mathcal{N}$, and component DNN modules of the cloud $z_{\zeta_{m}}$ and $u_{\lambda_{m}}$, $\forall m\in\mathcal{M}$. One naive approach is to adopt the standard stochastic SGD method, which simply updates a set of all trainable parameters $\theta=\varphi\bigcup\{\psi_{i}:\forall i\in\mathcal{N}\}$ simultaneously based on the gradient $\nabla_{\theta}l(\mathbf{x},\mathbf{t})$ obtained from the backpropagation algorithm.
To this end, the cloud should have access to the perfect knowledge of all local observation vectors. This poses a centralized data collection step invoking excessive fronthaul signaling overheads for distributed ENs. To address this difficulty, we propose a decentralized backpropagation algorithm where the ENs and cloud can update their own DNN parameters individually via the uplink-downlink fronthaul coordination.


\subsection{Cloud Update Strategy}
We first present an update policy of the cloud DNN, in particular, component DNNs $z_{\zeta_{m}}$ and $u_{\lambda_{m}}$, $\forall m\in\mathcal{M}$. As will be explained shortly, the proposed decentralized training algorithm resorts to uplink-downlink coordination at each training epoch. Thus, we can alternatively represent epoch as a fronthaul communication round. At the $k$-th communication round ($k=1,\cdots,K)$, the SGD update policies of $z_{\zeta_{m}}$ and $u_{\lambda_{m}}$ are given
\begin{subequations}\label{eq:cloud_update}
\begin{align}
\lambda_{m}^{[k]}&=\lambda_{m}^{[k-1]}-\eta\frac{1}{B}\sum_{b\in\mathcal{B}^{[k]}}\frac{\partial\mathbf{q}_{m}^{(b)}}{\partial\lambda_{m}^{[k-1]}}\nabla_{\mathbf{x}^{(b)}}l(\mathbf{x}^{(b)},\mathbf{t}^{(b)}),\\
\zeta_{m}^{[k]}&=\zeta_{m}^{[k-1]}-\eta\frac{1}{B}\sum_{b\in\mathcal{B}^{[k]}}\frac{\partial\mathbf{q}_{m}^{(b)}}{\partial\zeta_{m}^{[k-1]}}\nabla_{\mathbf{x}^{(b)}}l(\mathbf{x}^{(b)},\mathbf{t}^{(b)}),\label{eq:decoder_update}
\end{align}
\end{subequations}
where $V^{[k]}$ indicates the quantity of a variable $V$ at the $k$-th communication round, $\mathcal{B}^{[k]}$ is a set of mini-batch sample indices, $B\triangleq|\mathcal{B}^{[k]}|$ is the batch size, the superscript $b$ stands for the mini-batch sample index, and $\mathbf{x}^{(b)}$ and $\mathbf{t}^{(b)}$ respectively denote the output and label associated with the $b$-th local observation sample $\{\mathbf{a}_{i}^{(b)}:\forall i\in\mathcal{N}\}$. In \eqref{eq:cloud_update}, we have used the facts $\nabla_{\lambda_{m}}l(\mathbf{x},\mathbf{t})=\frac{\partial\mathbf{q}_{m}}{\partial\lambda_{m}}\nabla_{\mathbf{x}}l(\mathbf{x},\mathbf{t})$ and $\nabla_{\zeta_{m}}l(\mathbf{x},\mathbf{t})=\frac{\partial\mathbf{q}_{m}}{\partial\mathbf{\zeta}_{m}}\nabla_{\mathbf{x}}l(\mathbf{x},\mathbf{t})$. 

With a set of the labels $\{\mathbf{t}^{(b)}:\forall b\in\mathcal{B}^{[k]}\}$ at hands, the cloud readily obtains \eqref{eq:cloud_update} by means of the uplink fronthaul coordination from the ENs.
At the beginning of the $k$-th communication round, each EN $i$ sends the mini-batch set of the message vectors $\{\mathbf{s}_{i}^{(b)}:\forall b\in\mathcal{B}^{[k]}\}$ to the cloud through the corresponding uplink fronthaul channel $h_{i}$ \eqref{eq:yi_fading}. To convey $B$ mini-batch message vectors of length $S$, we need to assign $SB$ orthogonal RBs to each EN. Hence, the uplink fronthaul channel $h_{i}$, in particular, the channel matrix $\mathbf{H}_{i}$ and the Gaussian noise $\mathbf{n}_{C,i}$ in \eqref{eq:yi_fading}, generally varies for each mini-batch sample index $b$. To capture this effect, the uplink fronthaul channel model in \eqref{eq:yi_fading} is refined~to
\begin{align}\label{eq:yib}
    \mathbf{y}_{C,i}^{(b)}=\mathbf{H}_{i}^{(b)}\mathbf{s}_{i}^{(b)}+\mathbf{n}_{C,i}^{(b)}, 
\end{align} 
where $\mathbf{y}_{C,i}^{(b)}$ is the received signal at the cloud for the $b$-th mini-batch message $\mathbf{s}_{i}^{(b)}$ and $\mathbf{H}_{i}^{(b)}$ and $\mathbf{n}_{C,i}^{(b)}$ stand for the corresponding channel matrix and Gaussian noise, respectively.

Upon receiving the mini-batch message sets $\{\mathbf{s}_{i}^{(b)}:\forall b\in\mathcal{B}^{[k]}\}$, $\forall i\in\mathcal{N}$, the cloud performs the forward-pass computations in \eqref{eq:rmi}-\eqref{eq:x3} to obtain the outputs $\mathbf{x}^{(b)}$ and loss values $l(\mathbf{x}^{(b)},\mathbf{t}^{(b)})$, $\forall b\in\mathcal{B}^{[k]}$. Then, the backpropagation algorithm obtains the gradients $\nabla_{\lambda_{m}}l(\mathbf{x}^{(b)},\mathbf{t}^{(b)})$ and $\nabla_{\zeta_{m}}l(\mathbf{x}^{(b)},\mathbf{t}^{(b)})$ for all mini-batch samples $b\in\mathcal{B}^{[k]}$. To this end, the cloud first calculates the gradients of the loss function $\nabla_{\mathbf{x}^{(b)}}l(\mathbf{x}^{(b)},\mathbf{t}^{(b)})$, $\forall b\in\mathcal{B}^{[k]}$. Subsequently, the derivatives $\frac{\partial\mathbf{q}_{m}^{(b)}}{\partial\lambda_{m}^{[k-1]}}$ and $\frac{\partial\mathbf{q}_{m}^{(b)}}{\partial\zeta_{m}^{[k-1]}}$ can be attained individually. Combining these with $\nabla_{\mathbf{x}^{(b)}}l(\mathbf{x}^{(b)},\mathbf{t}^{(b)})$, we readily facilitates the SGD updates \eqref{eq:cloud_update} in parallel for each $\lambda_{m}$ and $\zeta_{m}$.

\subsection{EN Update Strategy}
After the cloud update process, individual ENs adjust their encoder parameters $\psi_{i}$, $\forall i\in\mathcal{N}$, in a decentralized manner. The associated gradient vector $\nabla_{\psi_{i}}l(\mathbf{x}^{(b)},\mathbf{t}^{(b)})$ of EN $i$ is derived as
\begin{align}\label{eq:grad_en}
\nabla_{\psi_{i}}l(\mathbf{x}^{(b)},\mathbf{t}^{(b)})=\frac{\partial\mathbf{s}_{i}^{(b)}}{\partial\psi_{i}}\mathbf{d}_{i}^{(b)}.
\end{align}
Here, the gradient vector $\mathbf{d}_{i}^{(b)}\in\mathbb{R}^{S}$ is defined as
\begin{subequations}
\begin{align}\label{eq:di}
\mathbf{d}_{i}^{(b)}&\triangleq
\frac{\partial\mathbf{y}_{C,i}^{(b)}}{\partial\mathbf{s}_{i}^{(b)}}\frac{\partial\mathbf{x}^{(b)}}{\partial\mathbf{y}_{C,i}^{(b)}}\nabla_{\mathbf{x}^{(b)}}l(\mathbf{x}^{(b)},\mathbf{t}^{(b)})\\
&=
\mathbf{H}_{i}^{(b)}\frac{\partial\mathbf{x}^{(b)}}{\partial\mathbf{y}_{C,i}^{(b)}}\nabla_{\mathbf{x}^{(b)}}l(\mathbf{x}^{(b)},\mathbf{t}^{(b)}),
\end{align}
\end{subequations}
where we have used the fact
$\frac{\partial\mathbf{y}_{C,i}^{(b)}}{\partial\mathbf{s}_{i}^{(b)}}=\mathbf{H}_{i}^{(b)}$.
Consequently, the SGD update strategy of $\psi_{i}$ is expressed as
\begin{align}\label{eq:en_update}
\psi_{i}^{[k]}=\psi_{i}^{[k-1]}-\eta\frac{1}{B}\sum_{b\in\mathcal{B}^{[k]}}\frac{\partial\mathbf{s}_{i}^{(b)}}{\partial\psi_{i}^{[k-1]}}\mathbf{d}_{i}^{(b)}.
\end{align}

Each EN $i$ can readily attain the derivative matrix $\frac{\partial\mathbf{s}_{i}^{(b)}}{\partial\psi_{i}^{[k-1]}}$ by employing the backpropagation algorithm locally through its encoder DNN $f_{\psi_{i}^{[k-1]}}$. On the contrary, the gradient vector $\mathbf{d}_{i}^{(b)}$ is not available at the ENs since they cannot access the gradient of the loss function $\nabla_{\mathbf{x}^{(b)}}l(\mathbf{x}^{(b)},\mathbf{t}^{(b)})$ straightforwardly. In fact, the cloud obtains this vector in advance for updating its DNN parameters \eqref{eq:cloud_update}. Thus, provided that the channel state information (CSI) $\{\mathbf{H}_{i}^{(b)}:\forall b\in\mathcal{B}^{[k]}\}$ is known, the cloud sends $\mathbf{d}_{i}^{(b)}$, $\forall b\in\mathcal{B}^{[k]}$, back to the corresponding ENs over reliable downlink fronthaul channels. Receiving the batch of the gradients $\{\mathbf{d}_{i}^{(b)}:\forall b\in\mathcal{B}^{[k]}\}$, each EN $i$ updates its encoder DNN parameter using \eqref{eq:en_update}.

\begin{algorithm}
\caption{Proposed Decentralized Training Algorithm}
\begin{algorithmic}[1]\label{alg:alg2}
\STATE{Set $k\leftarrow 0$.}
\STATE{EN $i$, $\forall i\in\mathcal{N}$, initializes $\psi_{i}[0]$.}
\STATE{Cloud initializes $\lambda_{m}[0]$ and $\zeta_{m}[0]$, $\forall m\in\mathcal{M}$.}
\STATE{Cloud and ENs predetermine mini-batch sets $\mathcal{B}^{[k]}$, $\forall k$.}
\FOR{communication round $k=1,2,\ldots,K$}
    \STATE{\textit{Edge forward-pass:}}
    \FOR{\textbf{each} EN $i\in\mathcal{N}$ \textbf{in parallel}}
        \STATE{EN $i$ calculates $\{\mathbf{s}_{i}^{(b)}:\forall b\}$ from \eqref{eq:si}.}
    \ENDFOR    
    \STATE{\textit{Uplink coordination:}}
    \FOR{\textbf{each} EN $i\in\mathcal{N}$ \textbf{in parallel}}
        \STATE{EN $i$ sends $\{\mathbf{s}_{i}^{(b)}:\forall b\}$ to cloud.}
    \ENDFOR  
    \STATE{\textit{Cloud backpropagation:}}
    \FOR{\textbf{each} $m\in\mathcal{M}$ \textbf{in parallel}}
        \STATE{Cloud updates $\lambda_{m}^{[k]}$ and $\zeta_{m}^{[k]}$, $\forall m\in\mathcal{M}$, from \eqref{eq:cloud_update}.}
    \ENDFOR  
    \STATE{Cloud computes $\{\mathbf{d}_{i}^{(b)}:\forall b\}$.}
    \STATE{\textit{Downlink coordination:}}
    \FOR{\textbf{each} EN $i\in\mathcal{N}$ \textbf{in parallel}}
        \STATE{Cloud sends $\{\mathbf{d}_{i}^{(b)}:\forall b\}$ to EN $i$.}
    \ENDFOR  
    \STATE{\textit{Edge backpropagation:}}
    \FOR{\textbf{each} EN $i\in\mathcal{N}$ \textbf{in parallel}}
        \STATE{EN $i$ updates $\psi_{i}^{[k]}$ from \eqref{eq:en_update}.}
    \ENDFOR
\ENDFOR
\end{algorithmic}
\end{algorithm}

\begin{figure}
\centering
\includegraphics[width=.7\linewidth]{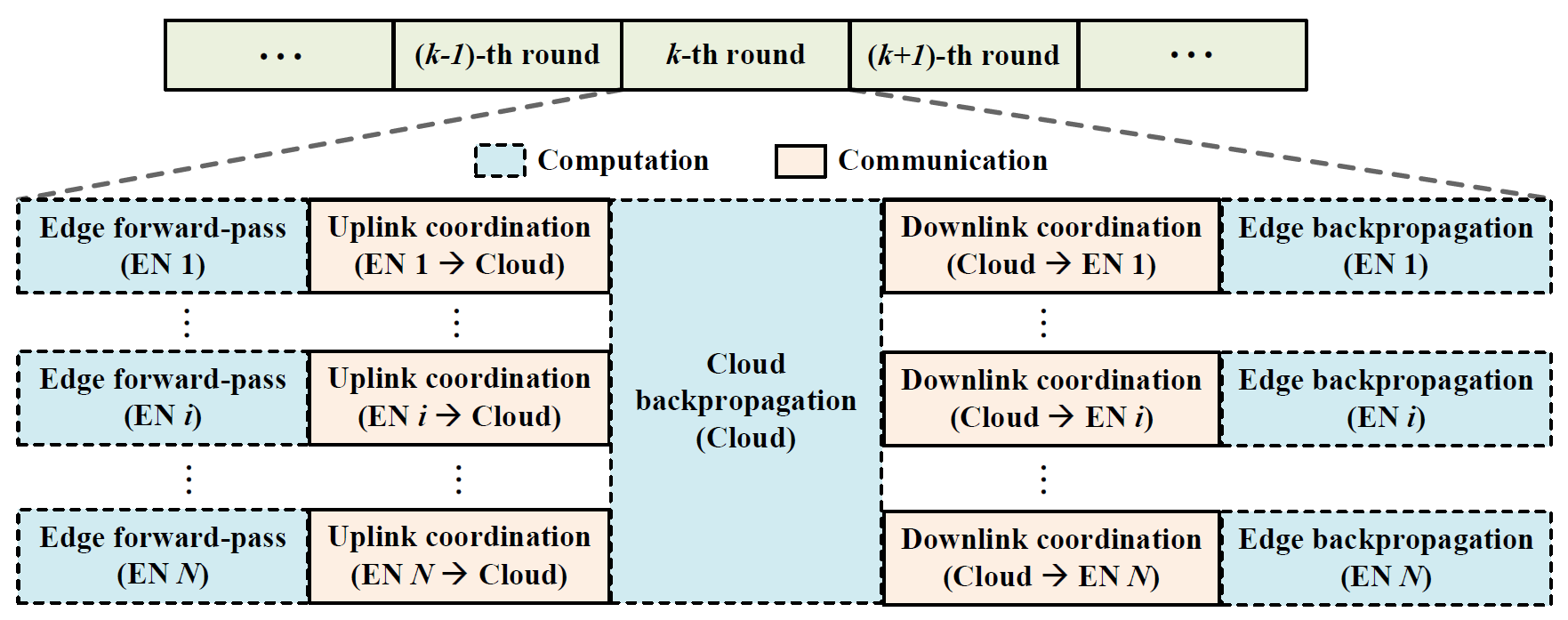}
\caption{Frame structure of one fronthaul communication round.}
\label{fig:fig3}
\end{figure}

Algorithm \ref{alg:alg2} summarizes the proposed decentralized training mechanism which optimizes the encoder DNNs and cloud DNN jointly over multiple fronthaul communication rounds. At the beginning of the training process, the ENs and cloud randomly generate a series of pseudo-random mini-batch sets $\mathcal{B}^{[k]}$ using predefined random seeds. As illustrated in Fig. \ref{fig:fig3}, each communication round consists of five sequential phases: edge forward-pass, uplink coordination, cloud backpropagation, donwlink coordination, and edge backpropagation.  
In the edge forward-pass phase, individual ENs compute their mini-batch message sets $\{\mathbf{s}_{i}^{(b)}:\forall b\in\mathcal{B}^{[k]}\}$ by leveraging the encoder DNN $f_{\psi_{i}}$. These are conveyed to the cloud in the subsequent uplink coordination phase. Receiving $\mathbf{y}_{C,i}^{(b)}$, $\forall b\in\mathcal{B}^{[k]}$, in the cloud backpropagation phase, the cloud employs the backpropagation algorithm to calculates the gradients required for the SGD updates in \eqref{eq:cloud_update}. In addition, the cloud obtains the mini-batch gradient set $\{\mathbf{d}_{i}^{(b)}:\forall b\in\mathcal{B}^{[k]}\}$, which is sent back to EN $i$ in the subsequent downlink coordination phase. 
It is then followed by the edge backpropagation phase where individual ENs perform the SGD updates in \eqref{eq:en_update} through the local backpropagation. Such an alternating update process of the cloud update and edge update is repeated until convergence.

The EN update rule in \eqref{eq:en_update} can be carried out individually without knowing the local observations of others. Also, the cloud update calculation \eqref{eq:cloud_update} only requests the mini-batch message sets $\{\mathbf{s}_{i}^{(b)}:\forall b\in\mathcal{B}^{[k]}\}$ sent by EN $i$, $\forall i\in\mathcal{N}$. Consequently, the proposed training algorithm can be implemented in a decentralized manner by exchanging fronthaul message $\mathbf{s}_{i}^{(b)}$ and gradient $\mathbf{d}_{i}^{(b)}$ in the uplink and downlink coordination phases, respectively. After the offline training, the optimized DNN parameters are installed at dedicated nodes for the real-time decentralized inference presented in Algorithm~\ref{alg:alg1}.

The number of ENs in the training step, denoted by $N_{train}$, would not perfectly match with that in the inference environment, denoted by $N_{test}$. Neverhteless, thanks to the versatile architecture, the cloud DNN trained with $N_{train}$ ENs can be directly applied to unmatched test environments consisting of $N_{test}\neq N_{train}$ ENs. Also, we can readily deploy $N_{train}$ trained encoder DNNs to the test edge network with smaller number of ENs $N_{test}\leq N_{train}$. Notice that $N_{train}$ is regarded as a hyperparameter of the proposed framework incurring a tradeoff relationship between the scalability and training performance. Increasing $N_{train}$ leads to a number of encoder DNNs that can be generalized to a wide range of the test EN populations. However, training with numerous ENs would be difficult as it invokes the optimization of $N_{train}$ encoder DNNs. Therefore, $N_{train}$ needs to be optimized carefully via the validation process. The impact of $N_{train}$ will be investigated using the numerical results.

\section{Extensions} \label{sec:sec5}
This section provides extension approaches of the proposed decentralized training algorithm to more practical scenarios.

\subsection{Wireless Backpropagation}

In the downlink coordination phase of Algorithm \ref{alg:alg2}, we assume an ideal noiseless downlink fronthaul links to share the gradient vectors $\mathbf{d}_{i}^{(b)}$, $\forall b\in\mathcal{B}^{[k]}$, to the ENs. In addition, for obtaining each $\mathbf{d}_{i}^{(b)}$, the cloud needs to know the perfect CSI $\mathbf{H}_{i}^{(b)}$, which would not practical. To address these difficulties, we present a wireless backpropagation policy where the gradients are successfully propagated through wireless downlink fronthaul links without the CSI knowledge at the cloud. As shown in Fig.~\ref{fig:fig3}, the proposed decentralized training strategy is built on sequential uplink-downlink fronthaul cooperation phases. This can be realized by the time division duplex (TDD) protocol where the channel reciprocity holds, i.e., the uplink and downlink fronthaul transmissions experience identical propagation environments. 

Based on this intuition, we provide an appropriate downlink coordination protocol that transfers the gradient vector $\mathbf{d}_{i}^{(b)}$ to EN $i$. Let $\mathbf{m}_{i}^{(b)}\in\mathbb{R}^{S}$ be a downlink fronthaul message sent from the cloud to EN~$i$, which bears the information about $\mathbf{d}_{i}^{(b)}$. The proposed approach constructs $\mathbf{m}_{i}^{(b)}$ as
\begin{align}\label{eq:mib}
    \mathbf{m}_{i}^{(b)}=\frac{\partial\mathbf{x}^{(b)}}{\partial\mathbf{y}_{C,i}^{(b)}}\nabla_{\mathbf{x}^{(b)}}l(\mathbf{x}^{(b)},\mathbf{t}^{(b)}).
\end{align}
Similar to \eqref{eq:sibb}, we partition $\mathbf{m}_{i}^{(b)}\in\mathbb{R}^{S}$ into real part $\mathbf{m}_{R,i}^{(b)}\in\mathbb{C}^{\tilde{S}}$ and imaginary part $\mathbf{m}_{I,i}^{(b)}\in\mathbb{C}^{\tilde{S}}$ with $\tilde{S}=S/2$. Then, the corresponding complex message $\tilde{\mathbf{m}}_{i}^{(b)}\in\mathbb{C}^{\tilde{S}}$ is obtained as $\tilde{\mathbf{m}}_{i}^{(b)}=\mathbf{m}_{R,i}^{(b)}+j\mathbf{m}_{I,i}^{(b)}$. To accommodate the transmit power budget $p_{C}$ at the cloud, we employ a power scaling factor $\alpha_{i}^{(b)}$ defined as
\begin{align}\label{eq:alpha}
    \alpha_{i}^{(b)}=\sqrt{\frac{p_{C}}{\max_{j}|\tilde{\mathbf{m}}_{i}^{(b)}[j]|^2}}.
\end{align}

Under the channel reciprocity, the baseband signal received at EN $i$, denoted by $\tilde{\mathbf{y}}_{E,i}^{(b)}\in\mathbb{R}^{\tilde{S}}$, is expressed as
\begin{align}    \tilde{\mathbf{y}}_{E,i}^{(b)}=\alpha_{i}^{(b)}(\tilde{\mathbf{h}}_{i}^{(b)})^{*}\odot\tilde{\mathbf{m}}_{i}^{(b)}+\tilde{\mathbf{n}}_{E,i}^{(b)},\label{eq:yei}
\end{align}
where $v^{*}$ stands for the complex conjugate of $v$, $(\tilde{\mathbf{h}}_{i}^{(b)})^{*}$ is the reciprocal downlink fronthaul channel vector for the $b$-th sample, and $\tilde{\mathbf{n}}_{E,i}^{(b)}\sim\mathcal{N}(\mathbf{0}_{\tilde{S}},\sigma_{E}^{2}\mathbf{I}_{\tilde{S}})\in\mathbb{R}^{\tilde{S}}$ stands for the zero-mean Gaussian noise at EN $i$ with variance~$\sigma_{E}^{2}$.

Likewise \eqref{eq:sip}, receiving $\tilde{\mathbf{y}}_{E,i}^{(b)}$, EN $i$ performs a linear decoding operation to compensate the phase error. The resulting decoded signal $\hat{\mathbf{y}}_{E,i}$ is given as
\begin{align}\label{eq:yEbar}
    \hat{\mathbf{y}}_{E,i}\triangleq e^{j\angle\tilde{\mathbf{h}}_{i}^{(b)}}\odot\frac{\tilde{\mathbf{y}}_{E,i}^{(b)}}{\alpha_{i}^{(b)}}
    =|\tilde{\mathbf{h}}_{i}^{(b)}|\odot\tilde{\mathbf{m}}_{i}^{(b)} + \hat{\mathbf{n}}_{E,i}^{(b)},
\end{align} 
where $\hat{\mathbf{n}}_{E,i}^{(b)}\triangleq e^{j\angle\tilde{\mathbf{h}}_{i}^{(b)}}\odot\frac{\tilde{\mathbf{n}}_{E,i}^{(b)}}{\alpha_{i}^{(b)}}\sim\mathcal{CN}(\mathbf{0}_{\tilde{S}},(\sigma_{E}/\alpha_{i}^{(b)})^2\mathbf{I}_{\tilde{S}})$.
For the real representation, we define $\mathbf{y}_{E,i}^{(b)}\triangleq[\Re\{\hat{\mathbf{y}}_{E,i}^{(b)}\}^{T},\Im\{\hat{\mathbf{y}}_{E,i}^{(b)}\}^{T}]^{T}$ and $\mathbf{n}_{E,i}^{(b)}\triangleq[\Re\{\hat{\mathbf{n}}_{E,i}^{(b)}\}^{T},\Im\{\hat{\mathbf{n}}_{E,i}^{(b)}\}^{T}]^{T}$.
Then, the effective transmission model from the cloud to EN $i$ becomes
\begin{align}
    \mathbf{y}_{E,i}^{(b)}
    =\mathbf{H}_{i}^{(b)}\mathbf{m}_{i}^{(b)}+\mathbf{n}_{E,i}^{(b)}
    =\mathbf{d}_{i}^{(b)}+\mathbf{n}_{E,i}^{(b)}, \label{eq:dib_noisy}
\end{align}
where the reciprocal downlink channel matrix $\mathbf{H}_{i}^{(b)}$ is consolidated into the gradient vector $\mathbf{d}_{i}^{(b)}$ as $\mathbf{d}_{i}^{(b)}=\mathbf{H}_{i}^{(b)}\mathbf{m}_{i}^{(b)}$. Therefore, the cloud does not need the CSI $\mathbf{H}_{i}^{(b)}$ to transfer the gradient vector $\mathbf{d}_{i}^{(b)}$ to EN $i$. 

The received signal vector $\mathbf{y}_{E,i}^{(b)}$ in \eqref{eq:dib_noisy} provides noisy gradient corrupted by the Gaussian noise $\mathbf{n}_{E,i}^{(b)}$. Nevertheless, it can be directly utilized to update the encoder DNN parameter $\psi_{i}$ by replacing $\mathbf{d}_{i}^{(b)}$ in \eqref{eq:en_update} with $\mathbf{y}_{E,i}^{(b)}$. This modified edge update rule is given as
\begin{align}\label{eq:en_update_wireless}
\psi_{i}^{[k]}=\psi_{i}^{[k-1]}-\eta\frac{1}{B}\sum_{b\in\mathcal{B}^{[k]}}\frac{\partial\mathbf{s}_{i}^{(b)}}{\partial\psi_{i}^{[k-1]}}\mathbf{y}_{E,i}^{(b)}.
\end{align}
The second term of \eqref{eq:en_update_wireless} can be rewritten by
\begin{subequations}
\begin{align}
    \frac{1}{B}\sum_{b\in\mathcal{B}^{[k]}}\frac{\partial\mathbf{s}_{i}^{(b)}}{\partial\psi_{i}^{[k-1]}}\mathbf{y}_{E,i}^{(b)}\label{eq:app}&=\frac{1}{B}\sum_{b\in\mathcal{B}^{[k]}}\frac{\partial\mathbf{s}_{i}^{(b)}}{\partial\psi_{i}^{[k-1]}}(\mathbf{d}_{i}^{(b)}+\mathbf{n}_{E,i}^{(b)})\\
    &\simeq\frac{1}{B}\sum_{b\in\mathcal{B}^{[k]}}\frac{\partial\mathbf{s}_{i}^{(b)}}{\partial\psi_{i}^{[k-1]}}\mathbf{d}_{i}^{(b)},\label{eq:grad_app}
\end{align}
\end{subequations}
where \eqref{eq:grad_app} comes from the facts that the Gaussian noise $\mathbf{n}_{E,i}^{(b)}$ has zero mean and is independent with $\mathbf{s}_{i}^{(b)}$ and $\mathbf{d}_{i}^{(b)}$. More precisely, for a sufficiently large batch size $B$, we have
\begin{subequations}\label{eq:ue}  
\begin{align}
\frac{1}{B}\sum_{b\in\mathcal{B}^{[k]}}\frac{\partial\mathbf{s}_{i}^{(b)}}{\partial\psi_{i}^{[k-1]}}\mathbf{n}_{E,i}^{(b)}
&\simeq\mathbb{E}\left[\frac{\partial\mathbf{s}_{i}}{\partial\psi_{i}^{[k-1]}}\mathbf{n}_{E,i}\right]\\
&=\mathbb{E}\left[\frac{\partial\mathbf{s}_{i}}{\partial\psi_{i}^{[k-1]}}\right]\mathbb{E}[\mathbf{n}_{E,i}]=\mathbf{0}_{S},
\end{align}
\end{subequations}
where the expectation is taken over the joint distribution of $\mathbf{s}_{i}$ and $\mathbf{n}_{E,i}$. It is inferred from \eqref{eq:ue} that \eqref{eq:app} becomes an unbiased estimate of the true gradient \eqref{eq:grad_app}. Therefore, for the wireless downlink fronthaul channels \eqref{eq:yei}, the ideal EN update policy in \eqref{eq:en_update}, which assumes noiseless downlink coordination, can be successfully modified into \eqref{eq:en_update_wireless}.

\begin{figure}
\centering
\includegraphics[width=.7\linewidth]{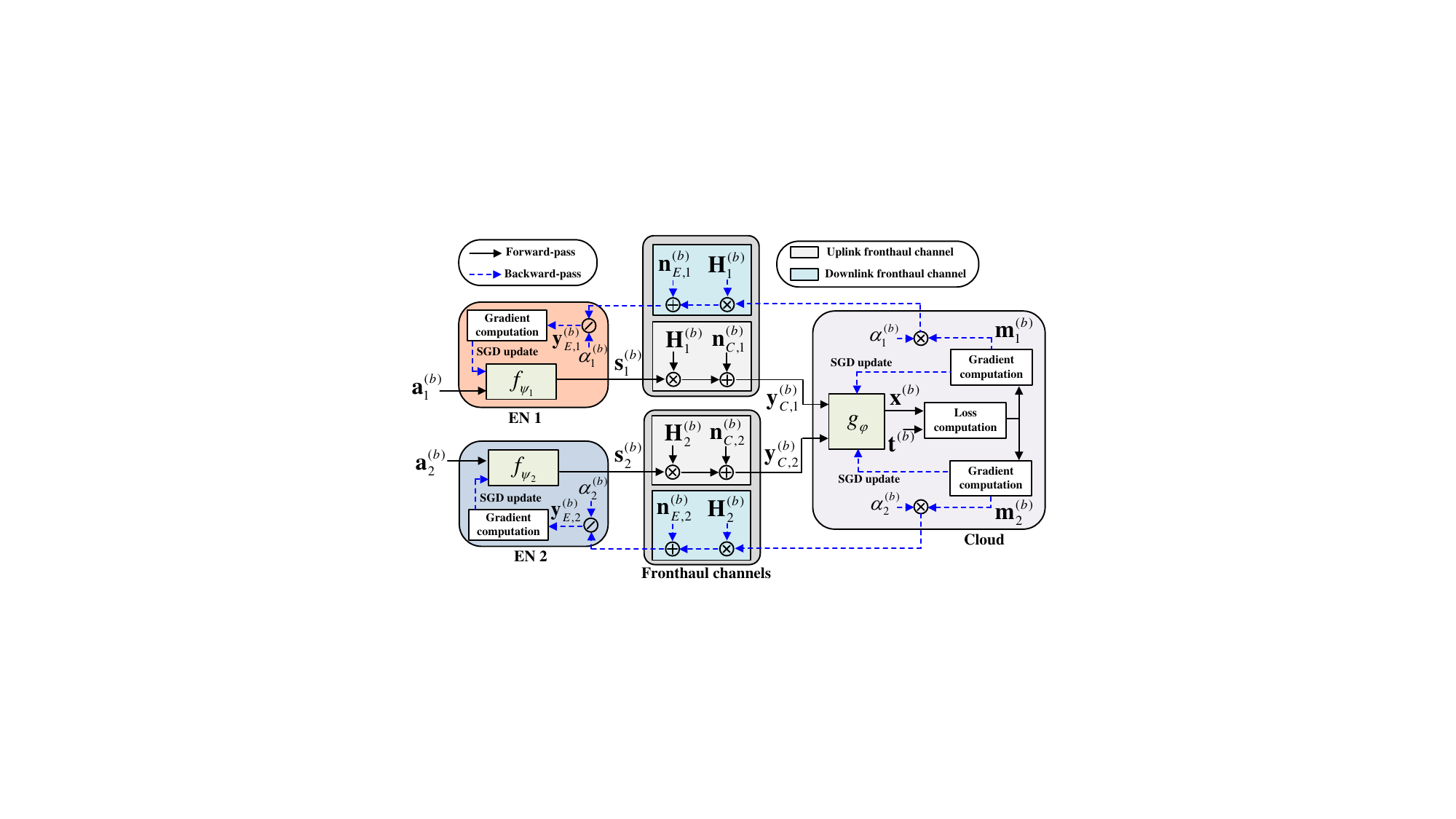}
\caption{Proposed wireless backpropagation mechanism with $N=2$ ENs.}
\label{fig:fig4}
\end{figure}

Fig. \ref{fig:fig4} illustrates the proposed wireless backpropagation mechanism, which can be easily included in Algorithm \ref{alg:alg2}. In the downlink coordination phase of Algorithm \ref{alg:alg2}, the cloud broadcasts the message vector $\tilde{\mathbf{m}}_{i}^{(b)}$ through the wireless downlink fronthaul channel \eqref{eq:yei}. Then, each EN $i$ recovers a noisy gradient vector $\mathbf{y}_{E,i}^{(b)}$ in \eqref{eq:dib_noisy} based on \eqref{eq:yEbar}. Accordingly, in the edge backpropagation phase of Algorithm \ref{alg:alg2}, the update rule of the ENs is replaced with \eqref{eq:en_update_wireless}. As a result, the proposed decentralized training algorithm can be implemented over the wireless fronthaul channels both in the uplink and downlink coordination phases. 

\subsection{Asynchronous Fronthaul Coordination}\label{sec:async}
So far, an ideal synchronous coordination case is assumed where all ENs can participate in the decentralized learning procedures. In practice, due to limited computing and communication resources, several ENs would not be share share their uplink fronthaul messages $\mathbf{s}_{i}^{(b)}$ for some mini-batch samples $b\in\mathcal{B}^{[k]}$. This invokes an asynchronous fronthaul coordination scenario where the number of active ENs stochastically changes for each mini-batch training sample of each communication round. Due to the absence of the uplink message $\mathbf{s}_{i}^{(b)}$, the cloud cannot calculate the associated downlink fronthaul message $\mathbf{m}_{i}^{(b)}$, thereby resulting in the failure of the entire training process.

To address this challenge, we modify the proposed decentralized training algorithm.  Let $\mathcal{N}^{[k]}(b)\subset[1,N_{train}]$ be the set of active ENs participating in the training for a particular local observation $\mathbf{a}_{i}^{(b)}$ at the $k$-th communication round. Also, we define $\mathcal{B}_{i}^{[k]}\subset\mathcal{B}^{[k]}$ as the index set of mini-batch samples that EN $i$ succeeds in sharing with the cloud at the $k$-th communication round. 
In the uplink coordination phase, the cloud can only receive $\mathbf{y}_{C,i}^{(b)}$, $\forall b\in\mathcal{B}_{i}^{[k]}$. Nevertheless, thanks to the versatile architecture of the cloud DNN, the estimate $\mathbf{x}^{(b)}$ can be obtained even though the number of the active ENs $|\mathcal{N}^{[k]}(b)|$ varies for each $b$. For the cloud DNN update \eqref{eq:decoder_update}, the derivative $\frac{\partial\mathbf{q}_{m}^{(b)}}{\partial\zeta_{m}^{[k-1]}}$ can be obtained as
\begin{align}
    \frac{\partial\mathbf{q}_{m}^{(b)}}{\partial\zeta_{m}^{[k-1]}}&=\left(\sum_{i\in\mathcal{N}^{[k]}(b)}\frac{\partial\mathbf{r}_{mi}^{(b)}}{\partial\zeta_{m}^{[k-1]}}\right)\frac{\partial\mathbf{q}_{m}^{(b)}}{\partial\mathbf{r}_{m}^{(b)}},
\end{align}
where we collect the derivatives $\frac{\partial\mathbf{r}_{mi}^{(b)}}{\partial\zeta_{m}^{[k-1]}}$ only for the active ENs $i\in\mathcal{N}^{[k]}(b)$. Likewise, the gradient vector $\mathbf{d}_{i}^{(b)}$ is sent only to the active EN $i\in\mathcal{N}^{[k]}(b)$ in the downlink coordination phase. In the subsequent edge backpropagation phase, the encoder DNN update in \eqref{eq:en_update_wireless} leverages the received signals $\mathbf{y}_{E,i}^{(b)}$ over active mini-batch samples $b\in\mathcal{B}_{i}^{[k]}$ as
\begin{align}\label{eq:en_update2}
\psi_{i}^{[k]}=\psi_{i}^{[k-1]}-\eta\frac{1}{|\mathcal{B}^{[k]}_{i}|}\sum_{b\in\mathcal{B}^{[k]}_{i}}\frac{\partial\mathbf{s}_{i}^{(b)}}{\partial\psi_{i}^{[k-1]}}\mathbf{y}_{E,i}^{(b)}.
\end{align}
By doing so, the proposed decentralized training algorithm can optimize the encoder and cloud DNNs successfully in the asynchronous fronthaul coordination case.

\subsection{Encoder Sharing Policy}\label{sec:encoder_sharing}


The proposed framework needs each EN $i$ to have the dedicated encoder DNN $f_{\psi_{i}}$. 
For this reason, the encoder DNNs trained at a certain training EN population $N_{train}$ cannot be straightforwardly deployed into a larger edge network with more ENs $N_{test}>N_{train}$. In addition, when the ENs share the identical knowledge basis and modality in their local input samples $\mathbf{a}_{i}$, $\forall i\in\mathcal{N}$, the dedicated encoders would not be viable for extracting useful features using partitioned input samples across the ENs. To address these difficulties, we exploit the parameter sharing technique in which the identical encoder DNN parameter $\psi$ is reused at all ENs. This imposes the following consensus constraint:
\begin{align}
    \psi_{i}=\psi,\ \forall i\in\mathcal{N}. \label{eq:cons}
\end{align}

We refine the training formulation in \eqref{eq:L} by imposing the encoder sharing constraint \eqref{eq:cons} as
\begin{subequations}\label{eq:cons_prob}
\begin{align}
    &\min_{\theta,\psi} L(\theta)\\
    &\text{subject to }\eqref{eq:cons}.
\end{align}
\end{subequations}
By fixing the cloud DNN parameter $\varphi$, problem \eqref{eq:cons_prob} boils down to the horizontal federated learning (HFL) task where the cloud manages the decentralized training of the shared encoder parameter $\psi$. A popular solution for this task is the federated averaging (FedAvg) algorithm~\cite{McMahan:17}. At the $k$-th communication round, individual ENs execute the decentralized SGD for their local encoder DNN parameters $\psi_{i}^{[k]}$, $\forall i\in\mathcal{N}$, from \eqref{eq:en_update}. It is then followed by the encoder aggregation at the cloud which builds the shared encoder parameter $\psi^{[k]}$~as
\begin{align}
    \psi^{[k]}=\frac{1}{N}\sum_{i\in\mathcal{N}}\psi_{i}^{[k]}. \label{eq:fedavg}
\end{align}
The shared encoder parameter $\psi^{[k]}$ is then dispatched to all ENs and it is utilized as an initial point of the local EN update. As a result, \eqref{eq:en_update} can be refined as
\begin{align}\label{eq:en_update_shared}
\psi_{i}^{[k]}=\psi^{[k-1]}-\eta\frac{1}{B}\sum_{b\in\mathcal{B}^{[k]}}\frac{\partial\mathbf{s}_{i}^{(b)}}{\partial\psi^{[k-1]}}\mathbf{d}_{i}^{(b)},
\end{align}
where, with a slight abuse of notations, $\mathbf{s}_{i}^{(b)}$ denotes the message of EN $i$ encoded by the shared encoder parameter $\psi$ , i.e., $\mathbf{s}_{i}^{(b)}=f_{\psi}(\mathbf{a}_{i}^{(b)})$.


To examine the viability of \eqref{eq:fedavg} and \eqref{eq:en_update_shared}, we investigate an ideal centralized training algorithm for solving \eqref{eq:cons_prob}. By substituting \eqref{eq:cons} into \eqref{eq:KA2}, the gradient $\nabla_{\psi}l(\mathbf{x},\mathbf{t})$ with respect to the shared encoder parameter $\varphi$ is derived as
\begin{align}
    \nabla_{\psi}l(\mathbf{x},\mathbf{t})&=\sum_{i\in\mathcal{N}}\frac{\partial\mathbf{s}_{i}}{\partial\psi}\frac{\partial\mathbf{y}_{C,i}}{\partial\mathbf{s}_{i}}\frac{\partial\mathbf{x}}{\partial\mathbf{y}_{C,i}}\nabla_{\mathbf{x}}l(\mathbf{x},\mathbf{t})=\sum_{i\in\mathcal{N}}\frac{\partial\mathbf{s}_{i}}{\partial\psi}\mathbf{d}_{i}.
\end{align}
The corresponding SGD update rule at the $k$-th communication round is given by
\begin{align}\label{eq:en_update_cent}
    \psi^{[k]}=\psi^{[k-1]}-\tilde{\eta}\sum_{i\in\mathcal{N}}\frac{1}{B}\sum_{b\in\mathcal{B}^{[k]}}\frac{\partial\mathbf{s}_{i}^{(b)}}{\partial\psi^{[k-1]}}\mathbf{d}_{i}^{(b)},
\end{align}
where $\tilde{\eta}$ indicates the learning rate.
Plugging \eqref{eq:en_update_shared} into \eqref{eq:fedavg} results~in
\begin{align}
    \psi^{[k]}=\psi^{[k-1]}-\frac{\eta}{N}\sum_{i\in\mathcal{N}}\frac{1}{B}\sum_{b\in\mathcal{B}^{[k]}}\frac{\partial\mathbf{s}_{i}^{(b)}}{\partial\psi^{[k-1]}}\mathbf{d}_{i}^{(b)}. \label{eq:en_update_decent}
\end{align}
We can see that \eqref{eq:en_update_decent} and \eqref{eq:en_update_cent} are equivalent with a scaling of the learning rate $\tilde{\eta}=\frac{\eta}{N}$. This implies that the centralized SGD rule \eqref{eq:en_update_cent} is alternatively carried out by the local EN update \eqref{eq:en_update_decent} and the encoder aggregation at the cloud \eqref{eq:fedavg}. 

The proposed encoding sharing policy can be readily included in Algorithm \ref{alg:alg2} together with the asynchronous fronthaul coordination method. The wireless backpropagation mechanism can also be employed in \eqref{eq:en_update_shared} by replacing the exact gradient $\mathbf{d}_{i}^{(b)}$ with the fronthaul-received signal $\mathbf{y}_{E,i}^{(b)}$ in \eqref{eq:dib_noisy}. The shared encoder DNN $f_{\psi}$ trained with $N_{train}$ ENs can be extended to larger edge networks $N_{test}>N_{train}$. 

\subsection{Sum Power Constraint}
So far, we consider the per-RB power constraint (PPC) where the uplink and downlink fronthaul message vectors $\mathbf{s}_{i}^{(b)}$ and $\mathbf{m}_{i}^{(b)}$ undergo element-wise magnitude constraints. This power constraint would be suitable for the frequency division multiple access scheme for restricting the radiated signal power of each frequency RB \cite{Palomar:03,Zhou:14,Sohrabi:17}. However, when it comes to the time division multiple access, the SPC would be a proper power constraint to measure the total power consumption for each transmission time block.

The proposed approach can be easily extended to the SPC. This can be achieved by modifying the output activation function of the edge encoder DNN \eqref{eq:proj} and the power scaling factor $\alpha_{i}^{(b)}$ for the wireless backpropagation \eqref{eq:alpha} as
\begin{subequations}
\begin{align}
    \mathbf{s}_{\sf{X},i}&=
    \begin{cases}
        \mathbf{v}_{\sf{X},i} & \text{if }\|\mathbf{v}_{i}\|^2 \leq p_{E}\\
        \sqrt{\frac{p_{E}}{q_i}}\mathbf{v}_{\sf{X},i} & \text{elsewhere}\label{eq:proj2}
    \end{cases}\\
    \alpha_{i}^{(b)}&=\sqrt{\frac{p_{C}}{\sum_{l\in\mathcal{N}}\|\tilde{\mathbf{m}}_{l}^{(b)}\|^{2}}}.
\end{align}
\end{subequations}

\subsection{Channel Quality Information at ENs}
The proposed framework works only with the channel direction information (CDI) $\angle\tilde{\mathbf{h}}_{i}$ at the ENs for the baseband signal processing \eqref{eq:ytilde} and \eqref{eq:yEbar}. When the ENs further have the channel quality information (CQI) $|\tilde{\mathbf{h}}_{i}|$, we can exploit this as a side input to the edge encoder DNN $f_{\psi_{i}}$ as
\begin{align}
    \mathbf{s}_{i}=f_{\psi_{i}}(\mathbf{a}_{i},|\tilde{\mathbf{h}}_{i}|). 
\end{align}
This approach, which is referred to as the proposed scheme with the CQI at the EN (CQIE), generates the channel-adaptive uplink fronthaul message $\mathbf{s}_{i}$. As a result, the edge encoder DNNs can learn task-oriented precoding strategies for improving the inference performance of the cloud.

\section{Relationships to Federated Learning}\label{sec:related}

The task-oriented edge network can be viewed as a generalization of the FL framework which trains DNN models using partitioned datasets of separate ENs with the help of the cloud. According to the dataset partitioning scenarios, the FL is classified into two different categories: HFL and VFL \cite{QYang:19}. ENs in the HFL systems are assumed to hold horizontally partitioned samples, i.e., different samples of the entire dataset. 
Thus, the major focus of the HFL is to train a shared DNN model that fits split datasets deployed at distributed ENs. 
On the contrary, in the VFL, ENs are assumed to hold vertically partitioned samples, i.e., each EN $i$ observes a local feature $\mathbf{a}_{i}$ of the full training data sample $\mathbf{a}$.
Thus, the proposed task-oriented edge network shares a similar design philosophy with the VFL framework. At the same time, we exploit the core idea of the HFL system, in particular, the FedAvg algorithm \cite{McMahan:17}, to develop the decentralized training strategy with the encoder sharing constraint. Nevertheless, fundamentals of the task-oriented edge network are, in general, different from the concepts and formalism of the HFL system. 

In the task-oriented edge network, it is crucial to develop a proper edge-cloud cooperative inference architecture. Nevertheless, such a model design issue has been abstracted in existing VFL methods. In \cite{YHu:19,JZhang:22,XZeng:22,YShi:23,LYang:22}, the cloud exploits a simple sum-aggregation inference rule given by 
\begin{align}
\mathbf{x}=\phi\left(\sum_{i\in\mathcal{N}}\mathbf{y}_{C,i}\right)=\phi\left(\sum_{i\in\mathcal{N}}h_{i}\circ f_{\psi_{i}}(\mathbf{a}_{i})\right), \label{eq:sp}
\end{align}
where $\phi$ indicates a fixed activation function, e.g., the softmax function for the classification task. The sum-aggregation structure \eqref{eq:sp} can be characterized as a learnable nomographic function in \eqref{eq:vim_ori}. The outer mapping $u_{\lambda}$ is simply set to the fixed activation $\phi$, indicating that there is no dedicated cloud DNN for processing the received signals $\mathbf{y}_{C,i}$, $\forall i\in\mathcal{N}$. Therefore, the role of the edge encoder DNNs $f_{\psi_{i}}$, $\forall i\in\mathcal{N}$, becomes more significant. This induces intensive neural calculations at the ENs to build very deep encoder architectures. In addition, \eqref{eq:sp} requires to match the dimension of the fronthaul message $\mathbf{s}_{i}=f_{\psi_{i}}(\mathbf{a}_{i})\in\mathbb{R}^{S}$ with the desired output $\mathbf{x}\in\mathbb{R}^{X}$. For this reason, the number of the RBs $S$ assigned to the uplink fronthaul coordination should be fixed as $S=X$, implying that proper fronthaul resource management is not viable.

Thus, a tractable approach is to employ the cloud DNN $g_{\varphi}$. One naive solution is to build a fully-connected neural network \cite{TChen:20,ZZhang:22,PLiu:22,Timothy:23} written by
\begin{align}
\mathbf{x}=g_{\varphi}(\mathbf{y}_{C}), \label{eq:cat}
\end{align}
which accepts the concatenation of the received signals $\mathbf{y}_{C}\triangleq[\mathbf{y}_{C,1}^{T},\cdots,\mathbf{y}_{C,N}^{T}]^{T}\in\mathbb{R}^{NS}$ as an input feature. To process the concatenated vector, the input dimension of the cloud DNN should scale with $N$, thereby resulting in a rigid architecture. For this reason, we need to prepare multiple cloud DNNs trained with all possible EN populations $N_{test}$ in the test environment.

A more flexible structure has been presented in \cite{CXie:22,JShao:23b,YShi:23b} which extends the sum-aggregation \eqref{eq:sp} by employing learnable heads $w_{\kappa_{i}}:\mathbb{R}^{S}\rightarrow\mathbb{R}^{X}$, $\forall i\in\mathcal{N}$, with $\kappa_{i}$ being the trainable parameter of the $i$-th head. This cloud DNN model is expressed as
\begin{align}
\mathbf{x}&=\phi\left(\sum_{i\in\mathcal{N}}w_{\kappa_{i}}\circ h_{i}\circ f_{\psi_{i}}(\mathbf{a}_{i}) \right), \label{eq:vim}
\end{align}
where the $i$-th head $w_{\kappa_{i}}$ is dedicated to handling the signal received from EN $i$. Unlike the sum-aggregation model which needs fixed message dimension $S=X$, \eqref{eq:vim} allow to adjust $S$ according to the number of fronthaul RBs. This cloud DNN model can also be interpreted as the nomographic representation \eqref{eq:capp}, which is not suitable to process continuous-valued signals $\mathbf{y}_{C,i}$, $\forall i\in\mathcal{N}$, received through the wireless fronthaul channels. Nevertheless, the sum-aggregation operation offers the versatile computation structure only for the test edge networks smaller than the training setup, i.e., $N_{test}\leq N_{train}$. 

Compared to these existing approaches, the proposed inference rule \eqref{eq:KA2} exploits the relationship between the VFL and the nomographic function. A careful investigation on the KA representation theorem \eqref{eq:KA} reveals that the oracle cloud inference can be constructed with two component DNNs $z_{\zeta_{m}}$ and $u_{\lambda_{m}}$. Unlike the multi-head sum-aggregation model \eqref{eq:vim}, the proposed cloud DNN in \eqref{eq:KA2} provides a valid inference model over the wireless fronthaul channels that generally produce continuous-valued channel outputs. In addition, the identical component DNNs $z_{\zeta_{m}}$ and $u_{\lambda_{m}}$ are straightforwardly adopted to process all the received signals $\mathbf{y}_{C,i}$, $\forall i\in\mathcal{N}$ universally. For this reason, a simple encoding sharing policy can bring the generalization capability for larger edge networks unseen during the training.

Another challenge is to involve the channel impairments in realizing DTDE learning strategies. Such an issue has been recently studied in \cite{XZeng:22} where the uplink message $\mathbf{s}_{i}$ is transferred to the cloud over wireless channels. For a simple sum-aggregation inference model \eqref{eq:sp}, receive filters at the cloud are optimized to minimize the mean-squared-error between the ground truth messages $\mathbf{s}_{i}$, $\forall i\in\mathcal{N}$, and their estimates. Scheduling policies of ENs have been presented in \cite{ZZhang:22} which determine active ENs participating in collaborative learning according to their channel conditions. 
These methods have focused only on the wireless uplink fronthaul channels in conveying $\mathbf{s}_{i}$ from each EN $i$ to the cloud. In contrast, the ideal noiseless downlink fronthaul links were assumed for sharing gradient vectors from the cloud to the ENs. The proposed wireless backpropagation protocol exploits the TDD fronthaul coordination to exchange both the uplink messages and the downlink gradient vectors through wireless fading channels. Moreover, the asynchronous coordination policy presented in Sec. \ref{sec:async} can inject arbitrary EN scheduling policies into the training process. As a result, the work in \cite{ZZhang:22} can be regarded as a special case of the proposed framework.
Recently, the VFL over the wireless uplink and downlink coordination channels has been investigated in \cite{YShi:23b}. By employing the conovex optimization techniques, toint optimization algorithms of uplink and downlink signal processing strategies were proposed. The design issue of the cloud inference model is ignored as this work assumed the sum-aggregation inference \eqref{eq:sp}. 

\section{Numerical Results} \label{sec:sec6}
Numerical results validating the proposed learning framework are presented. We consider image classification tasks of public datasets such as Tiny ImageNet \cite{TinyImageNet} and Food-101~\cite{FOOD101}. Tiny ImageNet dataset contains $120,000$ color images of 200 classes, each of which has $500$ training images, $50$ validation images, and $50$ test images. Food-101 dataset consists of $101,000$ color images with 101 classes, each of which has $750$ training images and $250$ test images. We resize the image size of both datasets into $3\times64\times64$, where the first dimension indicates color channels and the second and third dimensions represent the height and width, respectively. These image samples are regarded as the global network state $\mathbf{a}\in\mathbb{R}^{3\times64\times64}$. Each EN is assumed to observe a particular region of the global image $\mathbf{a}$ randomly cropped with window size $48\times48$, i.e., $\mathbf{a}_{i}\in\mathbb{R}^{3\times48\times48}$. 

\subsection{Implementation}

\begin{figure}
\centering
\includegraphics[width=.7\linewidth]{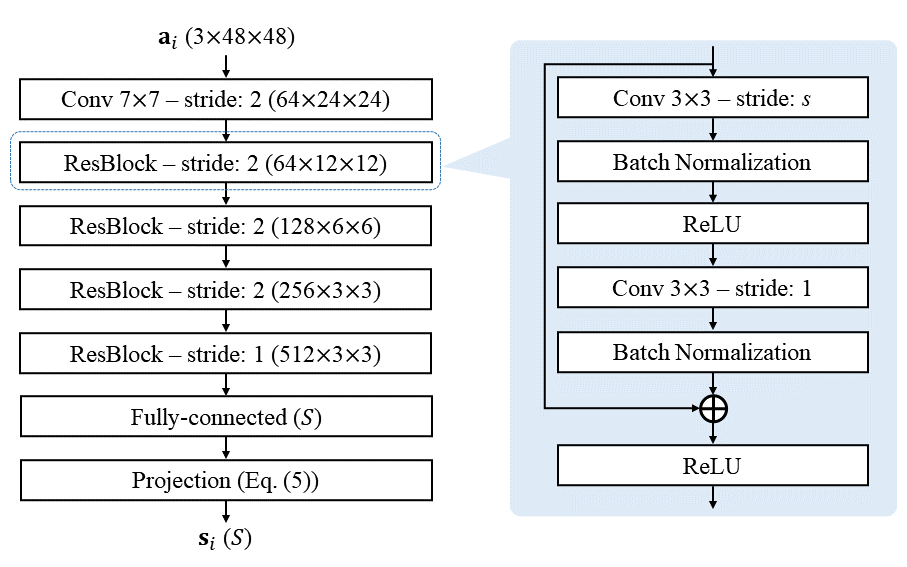}
\caption{Structure of encoder DNN. Output size of each layer is presented in parentheses.}
\label{fig:encoder}
\end{figure}

Fig.~\ref{fig:encoder} illustrates the structure of the encoder DNN $f_{\psi_{i}}$ which consists of one convolutional layer extracting low-level features of the local observation image $\mathbf{a}_{i}\in\mathbb{R}^{3\times48\times48}$, four residual blocks (ResBlocks) \cite{RESNET}, and one fully-connected layer that creates a message vector $\mathbf{s}_{i}\in\mathbb{R}^{S}$. The convolutional layer has $64$ kernels of size ${7}\times{7}$ with stride $2$, whose output size becomes $64\times 24\times 24$. The ResBlock comprises two convolutional layers with kernel size $3\times3$ and a skip connection link. In each ResBlock, we fix the stride of the second convolutional layer to $1$, whereas that of the first convolutional layer, denoted by $s$, is different such that $s=2$ and $1$ for the first three ResBlocks and the last ResBlock, respectively. Each ResBlock is designed to double its input channel dimension by adjusting the number of kernels of the first convolutional layers. The batch normalization is applied to the output of each convolutional layer along with the rectified linear unit (ReLU) activation. The skip connection link bypasses the input of the ResBlock to combine it with the output of the second batch normalization layer. The output of the fourth ResBlock of the encoder DNN is flattened into the vector processed by the fully-connected layer followed by the projection activation~\eqref{eq:proj}. The edge encoder DNN with the CQIE concatenates the channel vector $|\tilde{\mathbf{h}}_{i}|$ with the flattened output of the fourth ResBlock. The resulting concatenated vector is processed by the fully-connected layer.

The cloud DNN employs $M=17$ component modules $z_{\zeta_{m}}$ and $u_{\lambda_{m}}$ each of which is realized with two-layer MLP having $128$ neurons, resulting in about $8\times10^{5}$ trainable parameters. The ReLU activation is employed at hidden layers, whereas we apply the softmax activation at the output layer. We adopt the Adam optimizer~\cite{ADAM} with a learning rate $10^{-4}$. 

Unless stated otherwise, we consider the no CQIE scenario along with the PPC setup. The transmit power budgets at the ENs and the cloud are set to $p_{C}=p_{E}=1$. Also, we take into account the identical noise variances in the uplink and downlink fronthaul channels, i.e., $\sigma^{2}=\sigma_{C}^{2}=\sigma^{2}_{E}$. Then, the fronthaul signal-to-noise ratio (SNR) is defined as $\text{SNR}=1/\sigma^{2}$. In the training, we independently generate the uplink and downlink SNR values for each mini-batch sample uniformly within $[0\ \text{dB}, 30\ \text{dB}]$. By doing so, the resulting DNNs become adaptive to arbitrary changing propagation environment in the test step. The Rayleigh fading is considered for the fronthaul channel $\tilde{\mathbf{h}}_{i}^{(b)}$ which is independently generated for each sample $b$ and EN $i$ in the training, validation, and test processes. For the asynchronous fronthaul coordination, each EN is randomly dropped in the training step with the probability $\frac{({N}_{train}-1)}{2{N}_{train}}$. Thus, the average number of active ENs is slightly larger than half of the total ENs $N_{train}$.

\subsection{Evaluation of Proposed Decentralized Inference}
The proposed decentralized inference architecture is verified first. Unless stated otherwise, the batch size $B$ and the message dimension $S$ are fixed as $B=256$ and $S=16$, respectively. The proposed model is trained with $N_{train}=8$ ENs and its test accuracy is examined over a wider range of the test EN populations $N_{test}\in[4,8]$. The proposed wireless backpropagation mechanism is adopted for the decentralized training of encoder and cloud DNNs along with the asynchronous fronthaul coordination. No encoder sharing policy is employed in this subsection. For comparison, following three benchmark cloud DNN architectures are considered.
\begin{itemize}
    \item \textit{BaseNet}: The cloud is assumed to have the perfect access to the full image input $\mathbf{a}$. The cloud DNN consists of the encoder DNN in Fig. \ref{fig:encoder} except the final fully-connected layer. Instead, it is followed by a three-layer MLP with hidden dimension of 2048.
    \item \textit{MHNet}: The multi-head structure in \eqref{eq:vim} is adopted which has $N_{train}=8$ three-layer MLP heads $w_{\kappa_{i}}$, $\forall i\in[1,N_{train}]$. The output dimensions of hidden layers are fixed to 210.
    \item \textit{CatNet}: CatNet \eqref{eq:cat} is constructed as a three-layer MLP which accepts the concatenated received signal as an input feature. 
\end{itemize}
All the hidden layers are realized with the ReLU activation function. The depth and width of benchmark DNNs are set to preserve the similar model complexity with the proposed cloud DNN. BaseNet establishes an ideal task-oriented system where the cloud trivially knows the global image $\mathbf{a}$ with noiseless fronthaul coordination. The wireless backpropagation mechanism is applied for training MHNet and CatNet through practical fading fronthaul channels. The proposed cloud DNN and MHNet are trained at $N_{train}=8$ and are directly applied to the test EN population $N_{test}\in[4,8]$. Thus, they can be trained with the asynchronous fronthaul coordination scenario in Sec. \ref{sec:async}. 
On the contrary, the rigid structure of CatNet request perfectly matched training and test setups with $N_{train}=N_{test}$. For this reason, we assume the ideal synchronous coordination case in the training step of CatNet. The number of trainable parameters of CatNet scales with $N_{train}$ since its input dimension is given as $N_{train}S$. To keep the model complexity, we adjust the number of neurons of the hidden layers for each $N_{train}\in[4,8]$. 


\begin{figure}
\centering
    \subfigure[Tiny ImageNet]{
        \includegraphics[width=.45\linewidth]{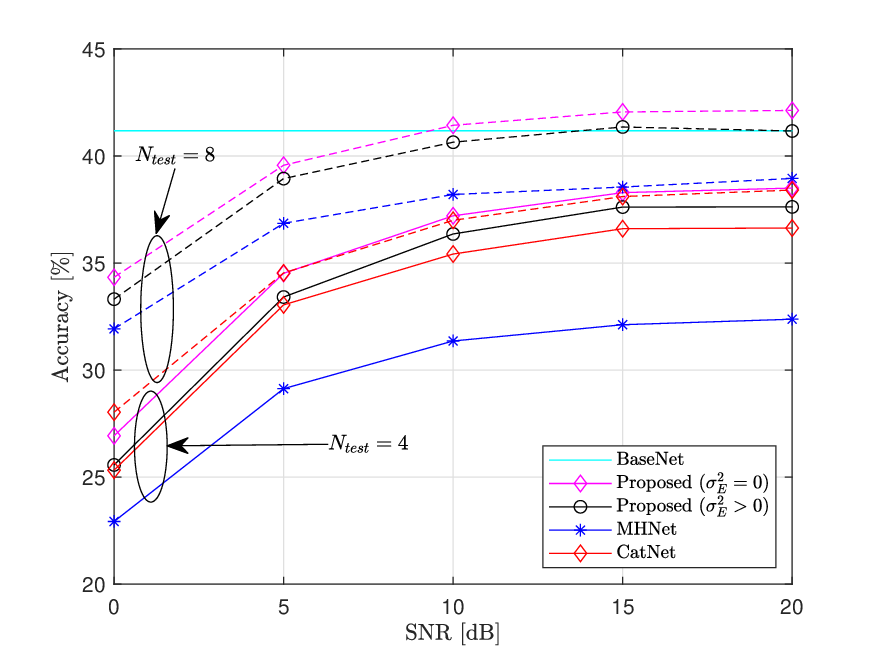}\label{fig:comp1a}
    }
    \subfigure[Food-101]{
        \includegraphics[width=.45\linewidth]{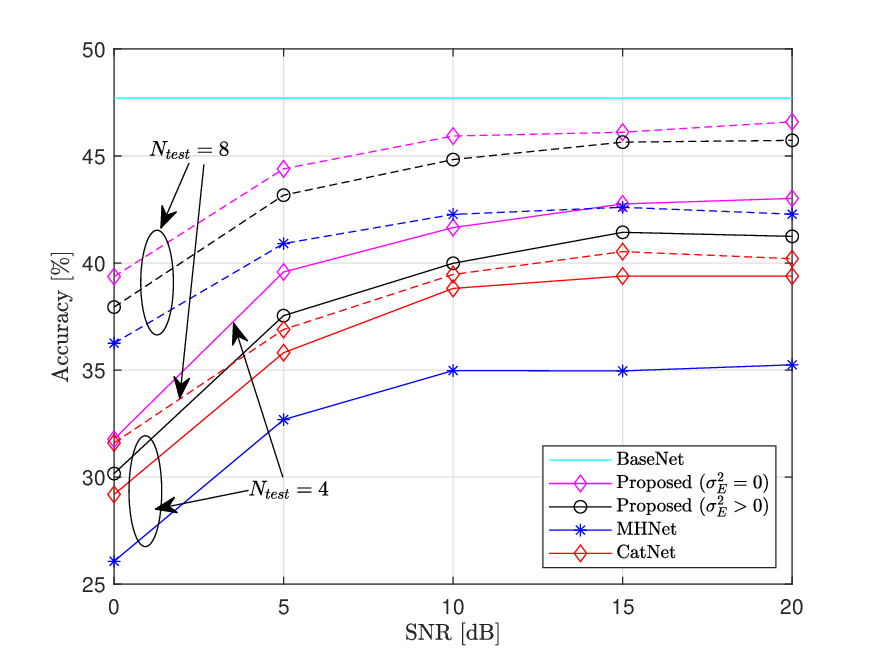}\label{fig:comp1b}
    }
    \caption{Accuracy performance with respect to SNR.}
    \label{fig:comp1}
\end{figure}

Fig. \ref{fig:comp1} presents the test accuracy performance of various methods with respect to the fronthaul SNR on Tiny ImageNet (Fig. \ref{fig:comp1a}) and Food-101 (Fig. \ref{fig:comp1b}). To verify the effectiveness of the wireless backpropagation strategy, we also plot the performance of the proposed cloud DNN that is trained over noise-free downlink fronthaul, i.e., $\sigma_{E}^{2}=0$. Regardless of the dataset and test EN populations $N_{test}$, the proposed approach outperforms MHNet and CatNet. As the SNR grows, the performance of the proposed scheme approaches the ideal performance generated by BaseNet. These results validate the effectiveness of the proposed cloud DNN architecture. The proposed method trained over noisy downlink fronthaul $\sigma_{E}^{2}>0$ achieves almost identical performance to that with $\sigma_{E}^{2}=0$. Therefore, we can conclude that the proposed wireless backpropagation mechanism successfully propagates valid gradient information through practical fading fronthaul channels. Increase in the test EN population $N_{test}$ improves the accuracy of all schemes. CatNet exhibits a good accuracy performance for the small number of ENs, i.e., $N_{test}=4$. However, when more ENs are deployed, its performance severely degrades compared to the proposed method and MHNet although the training and test environments of CatNet matches perfectly as $N_{train}=N_{test}$. This indicates that under the same model complexity, a simple concatenation-based cloud DNN model \eqref{eq:cat} would fail to estimate correct labels. As $N_{test}$ grows, CatNet generally requires more powerful computing architectures with additional neurons and layers to handle the concatenated input. The proposed scheme optimized for fixed EN population $N_{train}=8$ is superior to both CatNet and MHNet, proving the scalability of the proposed cloud DNN model.

\begin{figure}
\centering
    \subfigure[Tiny ImageNet]{
        \includegraphics[width=.45\linewidth]{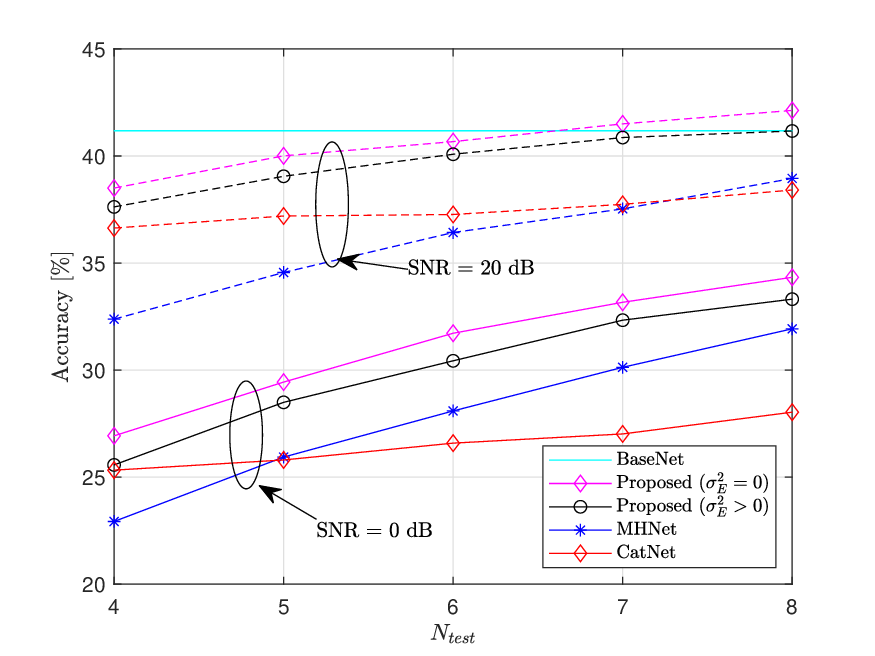}\label{fig:comp2a}
    }
    \subfigure[Food-101]{
        \includegraphics[width=.45\linewidth]{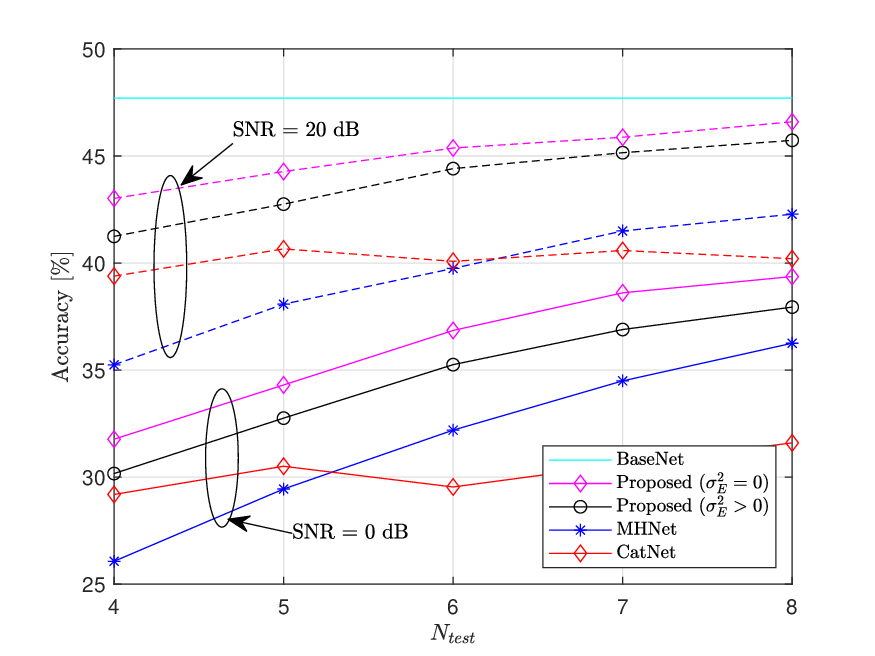}\label{fig:comp2b}
    }
    \caption{Accuracy performance with respect to $N_{test}$.}
    \label{fig:comp2}
\end{figure}

To see the scalability to the number of test ENs, we depict the accuracy performance in Fig. \ref{fig:comp2} by changing $N_{test}$. The proposed scheme trained with $\sigma_{E}^{2}>0$ shows negligible performance loss to the ideal noise-free downlink fronthaul case with $\sigma_{E}^{2}=0$. The proposed framework is superior to other baseline methods in all simulated $N_{test}$ and SNR. The accuracy of the proposed method is enhanced as $N_{test}$ grows. With $N_{test}=8$ ENs, it can achieve the upperbound performance of BaseNet on Tiny ImageNet dataset. This implies that the proposed cloud DNN architecture successfully leverages information sent by multiple ENs. On the contrary, the accuracy of CatNet does not improve with $N_{test}$, meaning that it fails to exploit the diversity of the local observations of the ENs. In the small $N_{test}$ regime, MHNet exhibits lower accuracy than CatNet. As discussed in Sec. \ref{sec:related}, the multi-head architecture \eqref{eq:vim} belongs to a simple nomographic representation \eqref{eq:capp}, which might not be valid for wireless fronthaul channels. Such a limitation can be addressed by the proposed cloud DNN facilitating the KA theorem \eqref{eq:KA} which provides a generic representation with arbitrary continuous-valued fronthaul channels. For this reason, the proposed approach performs better than MHNet regardless of the system parameters and datasets.

\begin{table}[]
\caption{Accuracy performance of various schemes for $N_{test}=4$}\label{tab:tab1}
\centering
\subtable[Tiny ImageNet]{
\begin{tabular}{c|cccccc|}
\cline{2-7}
                                 & \multicolumn{2}{c|}{$\text{SNR}=0\ \text{dB}$}            & \multicolumn{2}{c|}{$\text{SNR}=10\ \text{dB}$}           & \multicolumn{2}{c|}{$\text{SNR}=20\ \text{dB}$} \\ \hline
\multicolumn{1}{|c|}{Methods} & \multicolumn{1}{c|}{\!\!$S=16$\!\!} & \multicolumn{1}{c|}{\!\!$S=64$\!\!} & \multicolumn{1}{c|}{\!\!$S=16$\!\!} & \multicolumn{1}{c|}{\!\!$S=64$\!\!} & \multicolumn{1}{c|}{\!\!$S=16$\!\!}       & \!\!$S=64$\!\!      \\ \hline \hline
\multicolumn{1}{|c|}{Proposed}   & \multicolumn{1}{c|}{\textbf{25.57}}       & \multicolumn{1}{c|}{\textbf{36.18}}       & \multicolumn{1}{c|}{\textbf{36.36}}       & \multicolumn{1}{c|}{\textbf{41.91}}       & \multicolumn{1}{c|}{\textbf{37.62}}             &   \textbf{42.45}          \\ \hline
\multicolumn{1}{|c|}{MHNet}    & \multicolumn{1}{c|}{22.93}       & \multicolumn{1}{c|}{32.72}       & \multicolumn{1}{c|}{31.36}       & \multicolumn{1}{c|}{36.35}       & \multicolumn{1}{c|}{32.38}             &   36.96          \\ \hline
\multicolumn{1}{|c|}{CatNet}     & \multicolumn{1}{c|}{25.32}       & \multicolumn{1}{c|}{32.35}       & \multicolumn{1}{c|}{35.42}       & \multicolumn{1}{c|}{38.14}       & \multicolumn{1}{c|}{36.63}             &    38.75         \\ \hline
\end{tabular}
}
\subtable[Food-101]{
\begin{tabular}{c|cccccc|}
\cline{2-7}
                                 & \multicolumn{2}{c|}{$\text{SNR}=0\ \text{dB}$}            & \multicolumn{2}{c|}{$\text{SNR}=10\ \text{dB}$}           & \multicolumn{2}{c|}{$\text{SNR}=20\ \text{dB}$} \\ \hline
\multicolumn{1}{|c|}{Methods} & \multicolumn{1}{c|}{\!\!$S=16$\!\!} & \multicolumn{1}{c|}{\!\!$S=64$\!\!} & \multicolumn{1}{c|}{\!\!$S=16$\!\!} & \multicolumn{1}{c|}{\!\!$S=64$\!\!} & \multicolumn{1}{c|}{\!\!$S=16$\!\!}       & \!\!$S=64$\!\!      \\ \hline \hline
\multicolumn{1}{|c|}{Proposed}   & \multicolumn{1}{c|}{\textbf{37.27}}       & \multicolumn{1}{c|}{\textbf{44.94}}       & \multicolumn{1}{c|}{\textbf{45.80}}       & \multicolumn{1}{c|}{\textbf{48.36}}       & \multicolumn{1}{c|}{\textbf{46.00}}             &    \textbf{49.04}         \\ \hline
\multicolumn{1}{|c|}{MHNet}    & \multicolumn{1}{c|}{26.07}       & \multicolumn{1}{c|}{36.95}       & \multicolumn{1}{c|}{34.98}       & \multicolumn{1}{c|}{40.81}       & \multicolumn{1}{c|}{35.25}             &   41.24          \\ \hline
\multicolumn{1}{|c|}{CatNet}     & \multicolumn{1}{c|}{29.19}       & \multicolumn{1}{c|}{36.57}       & \multicolumn{1}{c|}{38.81}       & \multicolumn{1}{c|}{42.21}       & \multicolumn{1}{c|}{39.39}             &   43.06          \\ \hline
\end{tabular}
}
\end{table}

Table \ref{tab:tab1} presents the accuracy performance for $N_{test}=4$ with various combinations of message dimension $S$ and SNR. Here, the proposed scheme is trained over noisy downlink fronthaul links with $\sigma_{E}^{2}>0$. Boldface letters indicate the best accuracy performance. The performance of all schemes increases as the message dimension $S$ gets larger since the uplink messages become more informative. The proposed framework outperforms other baselines for all simulated setups, showing the effectiveness of the proposed cloud DNN.

\begin{table}[]
\caption{Accuracy performance of proposed scheme for various $M$ with $N_{train}=4$.}\label{tab:tab2}
\centering
\subtable[$\text{SNR}=0\ \text{dB}$]{
\begin{tabular}{c|cc|cc|}
\cline{2-5}
                          & \multicolumn{2}{c|}{$N_{test}=4$}                                                      & \multicolumn{2}{c|}{$N_{test}=8$}                                             \\ \hline
\multicolumn{1}{|c|}{$M$} & \multicolumn{1}{c|}{\!\!\!$S=16$\!\!\!}                          & \!\!\!$S=32$\!\!\!                          & \multicolumn{1}{c|}{\!\!\!$S=16$\!\!\!}                          & \!\!\!$S=32$\!\!\!               \\ \hline\hline
\multicolumn{1}{|c|}{5}   & \multicolumn{1}{c|}{25.51}                           & 31.82                           & \multicolumn{1}{c|}{33.24}                           & 38.82           \\ \hline
\multicolumn{1}{|c|}{9}   & \multicolumn{1}{c|}{\textbf{25.96}} & 31.98                           & \multicolumn{1}{c|}{\textbf{33.59}} & 39.00                           \\ \hline
\multicolumn{1}{|c|}{17}  & \multicolumn{1}{c|}{25.57}                           & \textbf{32.05} & \multicolumn{1}{c|}{33.31} & \textbf{39.25}                            \\ \hline
\end{tabular}
}
\subtable[$\text{SNR}=10\ \text{dB}$]{
\begin{tabular}{c|cc|cc|}
\cline{2-5}
                          & \multicolumn{2}{c|}{$N_{test}=4$}                                                      & \multicolumn{2}{c|}{$N_{test}=8$}                                             \\ \hline
\multicolumn{1}{|c|}{$M$} & \multicolumn{1}{c|}{\!\!\!$S=16$\!\!\!}                          & \!\!\!$S=32$\!\!\!                          & \multicolumn{1}{c|}{\!\!\!$S=16$\!\!\!}                          & \!\!\!$S=32$\!\!\!               \\ \hline\hline
\multicolumn{1}{|c|}{\!\!\!5\!\!\!}   & \multicolumn{1}{c|}{36.18}                           & 39.64                          & \multicolumn{1}{c|}{39.81}                           & 43.46           \\ \hline
\multicolumn{1}{|c|}{\!\!\!9\!\!\!}   & \multicolumn{1}{c|}{\textbf{36.79}} & 39.90                           & \multicolumn{1}{c|}{\textbf{40.70}} & 44.15                           \\ \hline
\multicolumn{1}{|c|}{\!\!\!17\!\!\!}  & \multicolumn{1}{c|}{36.36}                           & \textbf{40.34} & \multicolumn{1}{c|}{40.64} & \textbf{44.19}                            \\ \hline
\end{tabular}
}
\subtable[$\text{SNR}=20\ \text{dB}$]{
\begin{tabular}{c|cc|cc|}
\cline{2-5}
                          & \multicolumn{2}{c|}{$N_{test}=4$}                                                      & \multicolumn{2}{c|}{$N_{test}=8$}                                             \\ \hline
\multicolumn{1}{|c|}{$M$} & \multicolumn{1}{c|}{\!\!\!$S=16$\!\!\!}                          & \!\!\!$S=32$\!\!\!                          & \multicolumn{1}{c|}{\!\!\!$S=16$\!\!\!}                          & \!\!\!$S=32$\!\!\!               \\ \hline\hline
\multicolumn{1}{|c|}{5}   & \multicolumn{1}{c|}{37.03}                           & 40.28                           & \multicolumn{1}{c|}{40.61}                           & 43.87           \\ \hline
\multicolumn{1}{|c|}{9}   & \multicolumn{1}{c|}{\textbf{37.81}} & 40.94                           & \multicolumn{1}{c|}{\textbf{41.17}} & 44.61                           \\ \hline
\multicolumn{1}{|c|}{17}  & \multicolumn{1}{c|}{37.62}                           & \textbf{41.10} & \multicolumn{1}{c|}{\textbf{41.17}} & \textbf{44.68}                            \\ \hline
\end{tabular}
}
\end{table}

The impact of the number of component DNNs $M$ of the cloud DNN is examined in Table \ref{tab:tab2} on Tiny ImageNet dataset. Increasing $M$ leads to the improved expressive power of the cloud DNN model. For this reason, regardless of the simulation setups, the accuracy performance is generally enhanced at the expense of the inference complexity. For $S=16$, the proposed method with $M=9$ performs better than that with $M=17$, but the performance improvement is marginal. We thus choose $M=17$ for the rest of the simulations.

\begin{figure}
\centering
\includegraphics[width=.7\linewidth]{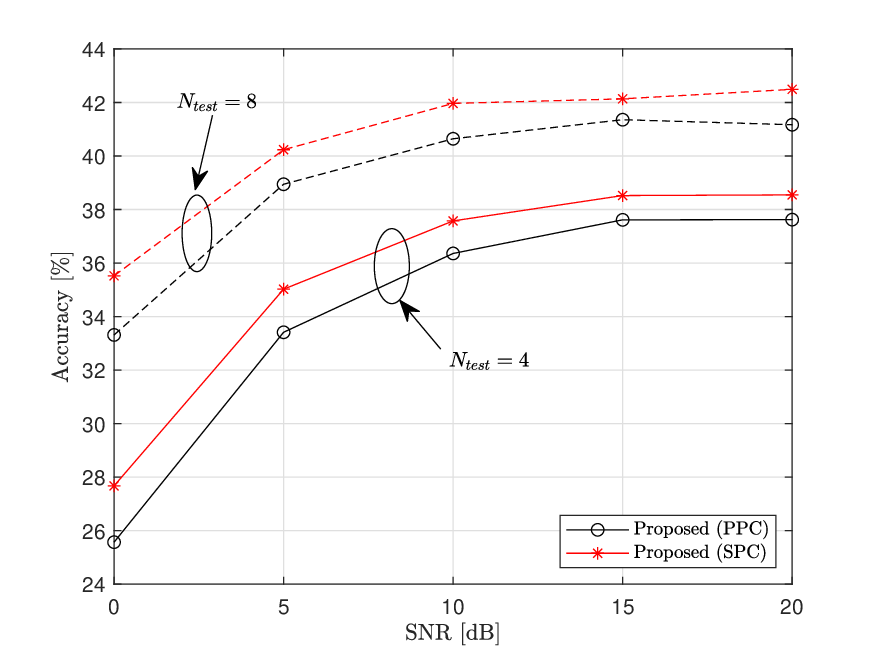}
\caption{Accuracy performance with respect to SNR for different power constraints.}
\label{fig:figB}
\end{figure}

Fig. \ref{fig:figB} depicts the accuracy performance of the proposed schemes with the PPC and SPC evaluated on the Tiny ImageNet dataset. For a fair comparison, in the SPC setup, the power budgets $p_{E}$ and $p_{C}$ at the ENs and cloud are set to $p_{E}=\tilde{S}$ and $p_{C}=N\tilde{S}$, respectively. Then, the PPC can be regarded as a special case of the SPC with equal power allocation. For this reason, the accuracy performance with the SPC performs better than that of the PPC. This indicates that the edge encoder DNNs control transmit powers of individual RBs to enhance the classification performance of the cloud DNN model.

\begin{table*}[]
\caption{Accuracy performance of proposed scheme with and without CQIE on Tiny ImageNet for $N_{test}=8$ and $S=16$.}\label{tab:tabA}
\centering
\subtable[w/o pathloss]{
\begin{tabular}{c|cccccc|}
\cline{2-7}
                                 & \multicolumn{2}{c|}{$\text{SNR}=0\ \text{dB}$}            & \multicolumn{2}{c|}{$\text{SNR}=10\ \text{dB}$}           & \multicolumn{2}{c|}{$\text{SNR}=20\ \text{dB}$} \\ \hline
\multicolumn{1}{|c|}{Methods} & \multicolumn{1}{c|}{\!\!$\sigma_{E}^{2}>0$\!\!} & \multicolumn{1}{c|}{\!\!$\sigma_{E}^{2}=0$\!\!} & \multicolumn{1}{c|}{\!\!$\sigma_{E}^{2}>0$\!\!} & \multicolumn{1}{c|}{\!\!$\sigma_{E}^{2}=0$\!\!} & \multicolumn{1}{c|}{\!\!$\sigma_{E}^{2}>0$\!\!}       & \!\!$\sigma_{E}^{2}=0$\!\!      \\ \hline \hline
\multicolumn{1}{|c|}{w/o CQIE}   & \multicolumn{1}{c|}{33.31}       & \multicolumn{1}{c|}{\textbf{34.33}}       & \multicolumn{1}{c|}{\textbf{40.64}}       & \multicolumn{1}{c|}{\textbf{41.43}}       & \multicolumn{1}{c|}{\textbf{41.17}}             &   \textbf{42.13}          \\ \hline
\multicolumn{1}{|c|}{w/ CQIE}    & \multicolumn{1}{c|}{\textbf{33.51}}       & \multicolumn{1}{c|}{34.30}       & \multicolumn{1}{c|}{40.33}       & \multicolumn{1}{c|}{40.75}       & \multicolumn{1}{c|}{40.88}             &   41.25          \\ \hline
\end{tabular}
}
\subtable[{w/ pathloss}]{
\begin{tabular}{c|cccccc|}
\cline{2-7}
                                 & \multicolumn{2}{c|}{$\text{SNR}=0\ \text{dB}$}            & \multicolumn{2}{c|}{$\text{SNR}=10\ \text{dB}$}           & \multicolumn{2}{c|}{$\text{SNR}=20\ \text{dB}$} \\ \hline
\multicolumn{1}{|c|}{Methods} & \multicolumn{1}{c|}{\!\!$\sigma_{E}^{2}>0$\!\!} & \multicolumn{1}{c|}{\!\!$\sigma_{E}^{2}=0$\!\!} & \multicolumn{1}{c|}{\!\!$\sigma_{E}^{2}>0$\!\!} & \multicolumn{1}{c|}{\!\!$\sigma_{E}^{2}=0$\!\!} & \multicolumn{1}{c|}{\!\!$\sigma_{E}^{2}>0$\!\!}       & \!\!$\sigma_{E}^{2}=0$\!\!      \\ \hline \hline
\multicolumn{1}{|c|}{w/o CQIE}   & \multicolumn{1}{c|}{0.57}       & \multicolumn{1}{c|}{0.58}       & \multicolumn{1}{c|}{0.66}       & \multicolumn{1}{c|}{0.69}       & \multicolumn{1}{c|}{0.92}             &         0.81    \\ \hline
\multicolumn{1}{|c|}{w/ CQIE}    & \multicolumn{1}{c|}{\textbf{28.54}}       & \multicolumn{1}{c|}{\textbf{29.81}}       & \multicolumn{1}{c|}{\textbf{30.06}}       & \multicolumn{1}{c|}{\textbf{31.75}}       & \multicolumn{1}{c|}{\textbf{30.21}}             &   \textbf{31.80}          \\ \hline
\end{tabular}
}
\end{table*}

Table \ref{tab:tabA} investigates the importance of the CQIE $|\mathbf{h}_{i}|$ in designing the proposed task-oriented edge networks. We consider two different channel models according to the presence of the pathloss. The distance-based pathloss model is employed where the complex channel vector $\tilde{\mathbf{h}}_{i}$ is generated as $\tilde{\mathbf{h}}_{i}\sim\mathcal{CN}(\mathbf{0}_{\tilde{S}},d_{i}^{-\alpha}\mathbf{I}_{\tilde{S}})$. Here, $d_{i}$ stands for the distance between EN $i$ and cloud uniformly distributed within $[10,50]$ m and $\alpha=2.7$ is the pathloss exponent. Table \ref{tab:tabA}(a) exhibits the accuracy performance of the proposed method without the pathloss. Exploiting the CQI as the additional input feature has no critical impact on the accuracy regardless of the SNR and the existence of the downlink fronthaul noise. Thus, we can conclude that simple matched-filtering processes in \eqref{eq:ytilde} and \eqref{eq:yEbar} achieve good performance by mitigating the phase ambiguity. As we can see from Table \ref{tab:tabA}(b), in the presence of the pathloss, the accuracy performance is severely degraded if the ENs do not use the CQI. Only with the CDI knowledge, it is highly difficult to compensate for the randomness in the message amplitude induced by the pathloss. Thus, it requires a proper precoding strategy that can mitigate the heterogeneous channel gains stemming from the random pathloss. This can be resolved by using the CQI knowledge at the edge encoder DNNs. The accuracy with the CQIE significantly increases the accuracy performance both in the ideal noiseless downlink fronthaul $\sigma_{E}^{2}=0$ and practical noisy fronthaul $\sigma_{E}^{2}>0$. This implies that the channel state information at the ENs plays a critical role in realizing practical task-oriented edge networks.

\subsection{Evaluation of Proposed Decentralized Training}

\begin{figure}
\centering
\includegraphics[width=.7\linewidth]{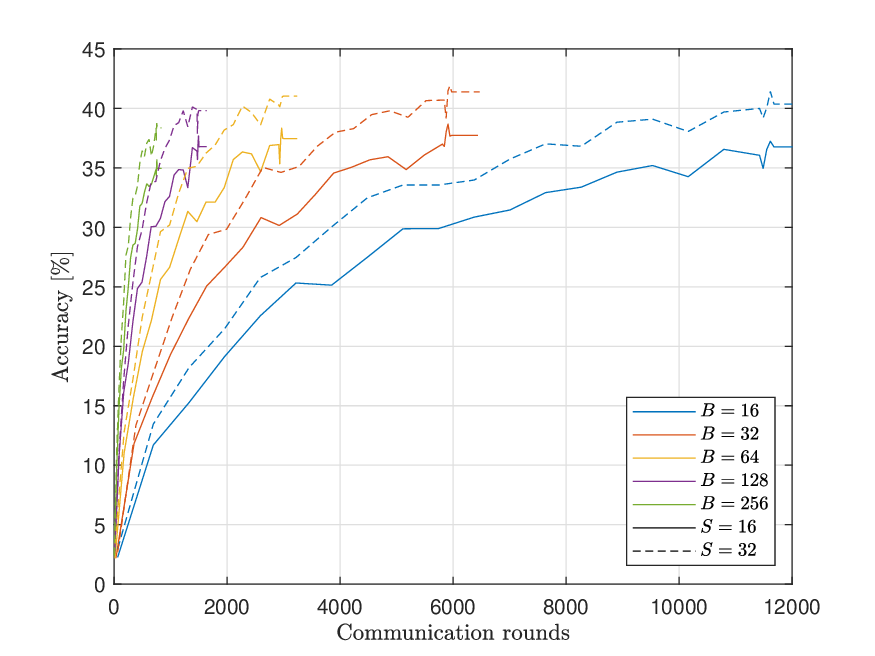}
\caption{Convergence behavior of proposed decentralized training algorithm for $N_{train}=4$.}
\label{fig:secVI_B_fig1}
\end{figure}

\begin{table}
\caption{Communication rounds to achieve accuracy of $35\%$}\label{tab:tab3}
\centering
\begin{tabular}{c|c|c|c|c|c|}
\cline{2-6}
                            & \multicolumn{1}{l|}{$B=16$} & \multicolumn{1}{l|}{$B=32$} & \multicolumn{1}{l|}{$B=64$} & \multicolumn{1}{l|}{$B=128$} & \multicolumn{1}{l|}{$B=256$} \\ \hline
\multicolumn{1}{|l|}{$S=16$} & 9529                        & 4206                        & 2110                        & 1386                         & \textbf{753}                        \\ \hline
\multicolumn{1}{|l|}{$S=32$} & 7005                        & 2633                        & 1466                        & 819                        & \textbf{456}                        \\ \hline
\end{tabular}
\end{table}

Next, we assess the proposed decentralized training scheme on Tiny ImageNet dataset with various hyperparameter setups. The batch size determines the accuracy of the approximation \eqref{eq:grad_app} in the wireless backpropagation process. For a large $B$, the additive noise in the encoder DNN update can be successfully mitigated, and thus we can expect the enhanced training performance. 
Such an issue is investigated in Fig.~\ref{fig:secVI_B_fig1} which exhibits the convergence behavior of the proposed decentralized training strategy by depicting the validation accuracy in terms of the communication rounds. Solid and dashed lines indicate the performance of the proposed scheme with message dimensions $S=16$ and $32$, respectively. We also mark the benchmark accuracy performance of $35\ \%$ with magenta solid line, where the number of the communication rounds for achieving this performance is summarized in Table \ref{tab:tab3}. As expected, increasing $B$ leads to faster convergence in terms of the communication rounds. As a result, the training latency for achieving the accuracy of $35\ \%$ can be minimized by adopting large $B$. The message dimension $S$ also affects the convergence speed as well as the test accuracy performance presented in Table \ref{tab:tab1}. It is noted that the number of RBs required for uplink and downlink fronthaul coordination is given by $BS$. Thus, such a performance improvement needs additional communication cost at each round.

\begin{figure}
\centering
    \subfigure[$S=16$, $\text{SNR}=0\ \text{dB}$]{
        \includegraphics[width=0.3\linewidth]{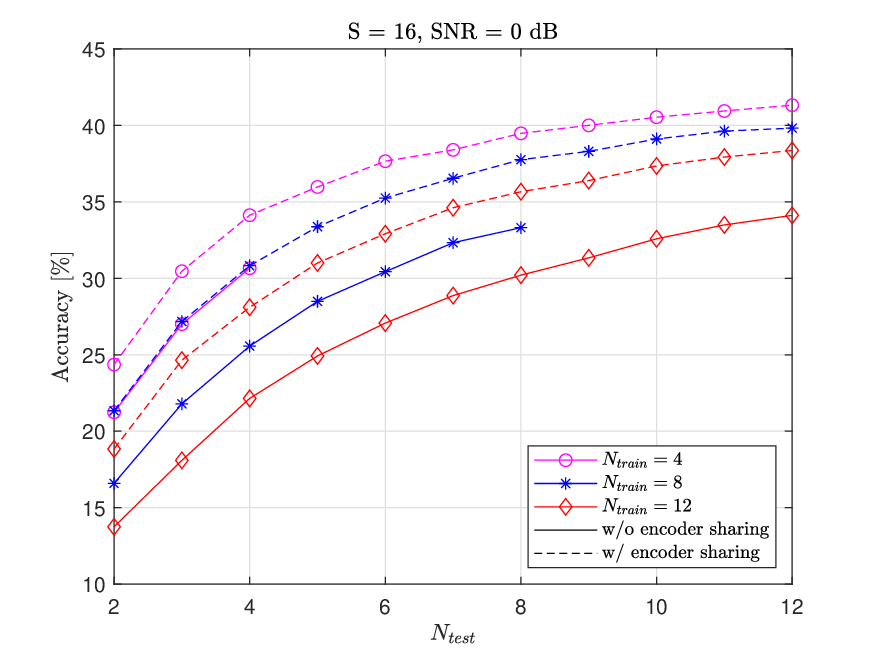}\label{fig:simul2a}
    }\hspace{-7mm}
        \subfigure[$S=32$, $\text{SNR}=0\ \text{dB}$]{
        \includegraphics[width=0.3\linewidth]{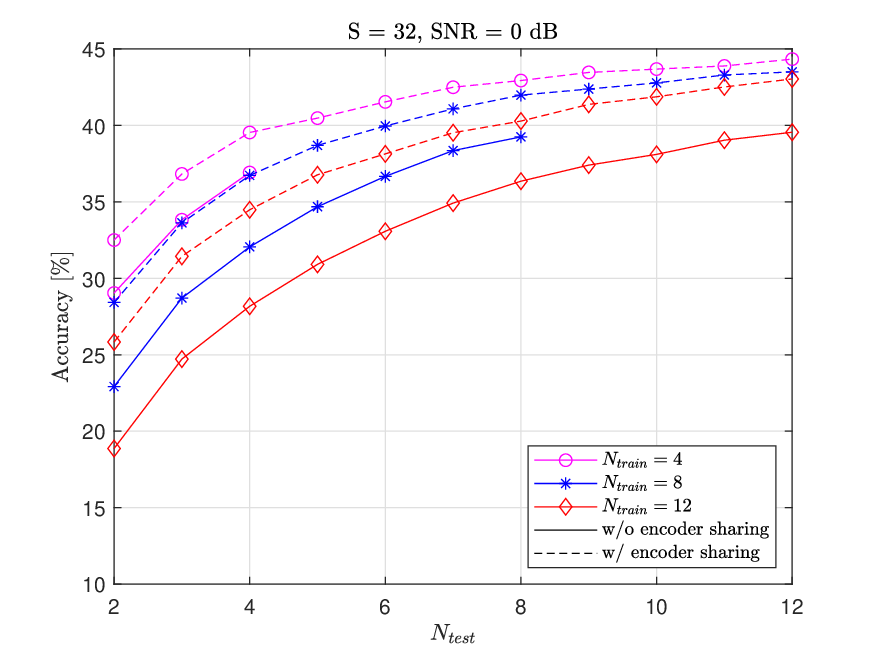}\label{fig:simul2b}
    }\hspace{-7mm}
    \subfigure[$S=64$, $\text{SNR}=0\ \text{dB}$]{
        \includegraphics[width=0.3\linewidth]{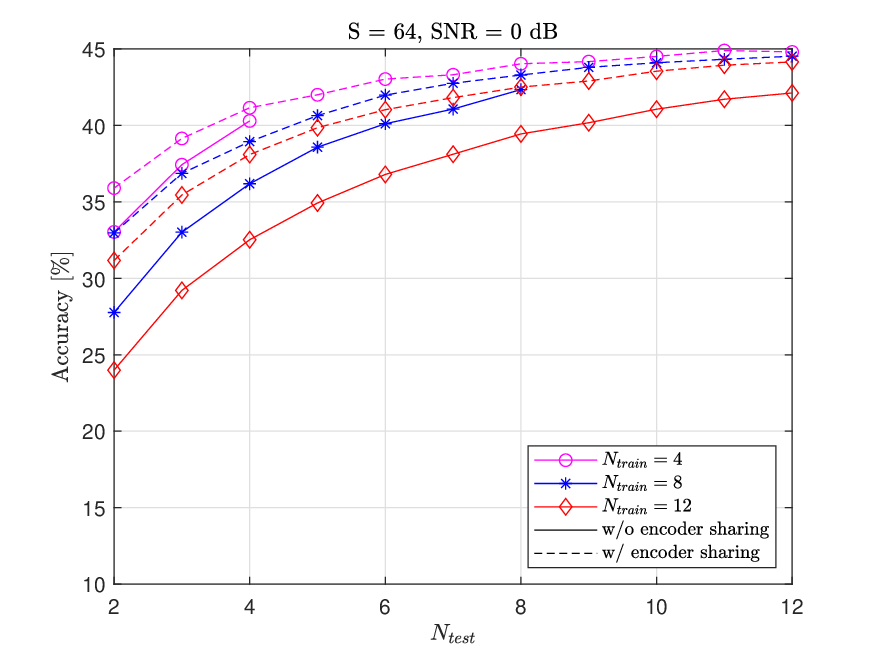}\label{fig:simul2c}
    }
    \subfigure[$S=16$, $\text{SNR}=20\ \text{dB}$]{
        \includegraphics[width=0.3\linewidth]{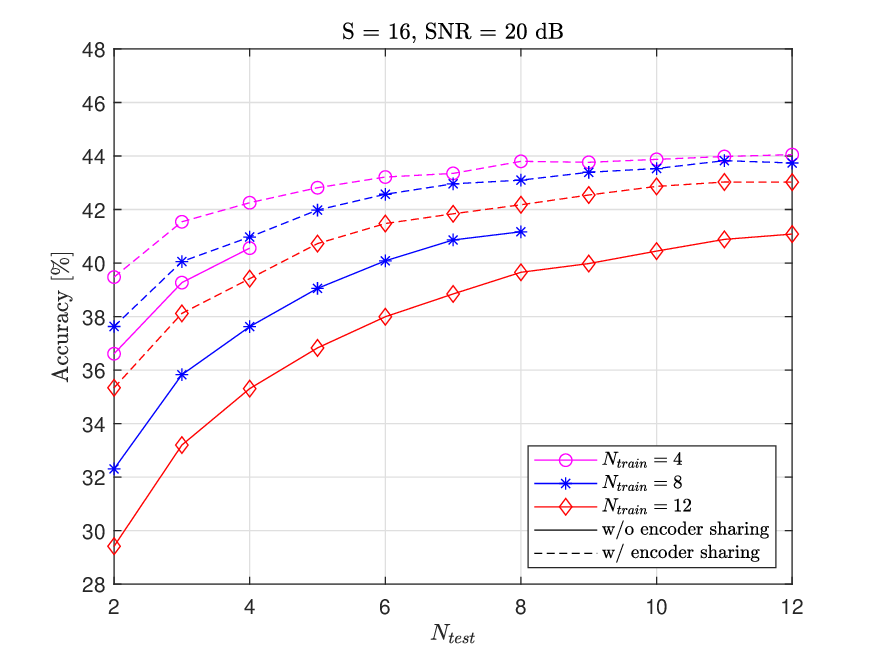}\label{fig:simul2d}
    }\hspace{-7mm}
        \subfigure[$S=32$, $\text{SNR}=20\ \text{dB}$]{
        \includegraphics[width=0.3\linewidth]{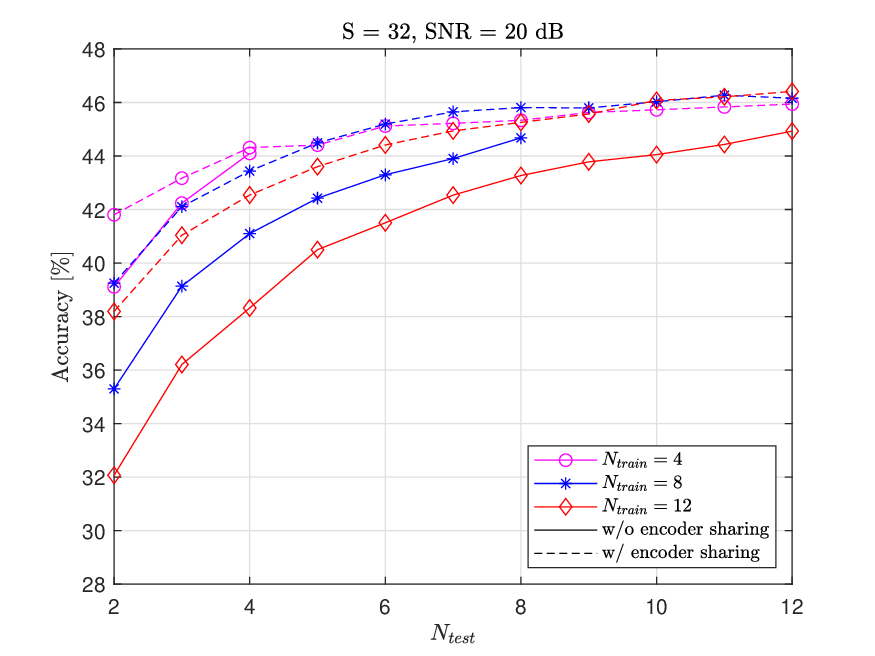}\label{fig:simul2e}
    }\hspace{-7mm}
    \subfigure[$S=64$, $\text{SNR}=20\ \text{dB}$]{
        \includegraphics[width=0.3\linewidth]{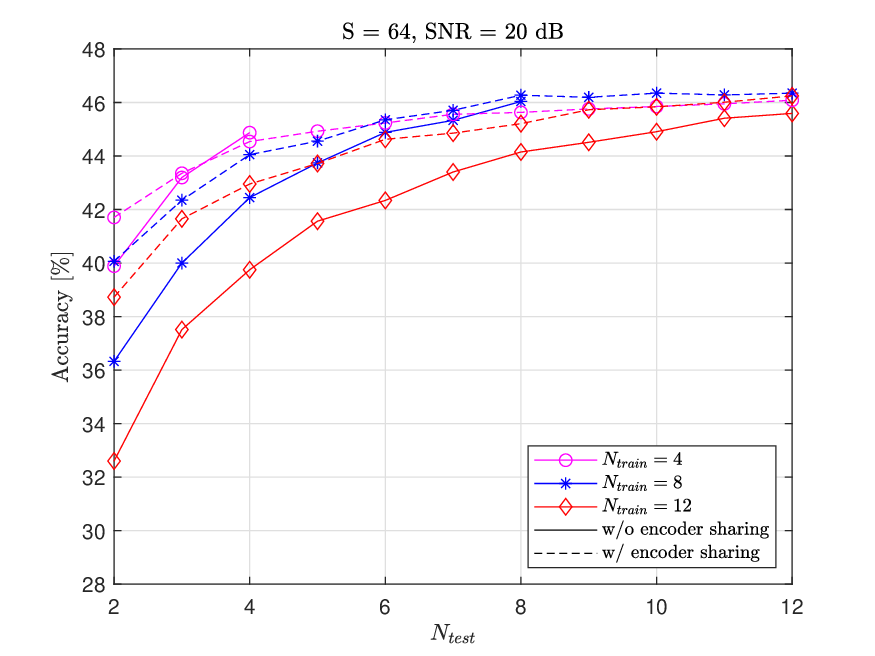}\label{fig:simul2f}
    }
    \caption{Accuracy performance with respect to $N_{test}$. Results in each row and column are obtained for different SNR and message dimensions $S$, respectively.}
    \label{fig:secVI_B_fig2}
\end{figure}

Fig. \ref{fig:secVI_B_fig2} validates the proposed encoder sharing policy by evaluating the accuracy performance by changing $N_{test}\in[2,12]$. The first and second rows exhibit the results obtained for $\text{SNR}=0$ and $20$ dB, respectively. Also, each column corresponds to different message dimensions $S\in\{16,32,64\}$. Solid and dashed lines respectively indicate the proposed training algorithm without and with the encoder sharing policy. Without sharing the encoder DNNs, the proposed method is valid only for $N_{test}\leq N_{train}$. On the contrary, the encoder sharing policy directly reuses a sole encoder DNN for all test EN populations $N_{test}\in[2,12]$. We can see that the proposed encoder sharing policy improves the accuracy performance regardless of the SNR and message dimension $S$. Such a performance improvement is obtained by allowing the shared encoder DNN to observe a number of local image inputs $\mathbf{a}_{i}$ captured by all ENs $i\in\mathcal{N}$. By doing so, we can improve the generalization ability of the proposed framework for unseen EN populations, i.e, $N_{test}>N_{train}$. The encoder sharing policy becomes more powerful in severe communication environments with low SNR and small $S$ regimes where the cloud might fail to get informative messages from the ENs due to the large channel noise and high compression rate. The proposed framework with the encoder sharing policy trained at $N_{train}=4$ is superior to those trained at other settings. Thus, we can conclude that few ENs are sufficient to achieve a good accuracy performance in a wide range of the test EN populations $N_{test}\in[2, 12]$. A similar phenomenon can be observed for the cases without the encoder sharing policy where the proposed method trained at $N_{train}=4$ outperforms other setups. However, such a small number of the training ENs results in a poor scalability as it can only be employed for $N_{test}\leq N_{train}$. This validates the effectiveness of the proposed encoder sharing policy that enhances the accuracy performance and the scalability simultaneously.

\section{Conclusions} \label{sec:sec7}

This paper has investigated the DTDE policy for optimizing task-oriented edge networks. Collaboration among the ENs and cloud is available only through resource-constrained wireless fronthaul links. This requires properly designed uplink-downlink fronthaul coordination protocols to facilitate decentralized inference and training. To this end, we have first developed a cooperative inference in the presence of wireless uplink fronthaul links. Inspired by the nomographic function, the oracle inference architecture can be built using a group of individual edge encoder DNNs and a cloud DNN having multi-branch component DNNs. This approach offers a versatile computation structure of the cloud DNN that is independent of the number of ENs. A decentralized training strategy of the ENs and cloud has been proposed where encoder and cloud DNNs are optimized by exchanging the gradient information from the cloud to ENs via wireless downlink fronthaul channels. Several extension approaches of the proposed framework to more practical coordination scenarios have also been presented. We have demonstrated the effectiveness of the proposed framework for image classification tasks. Numerical results have validated the superiority of the proposed approach over existing methods. The proposed framework requires orthogonal RBs for multiple ENs in realizing uplink and downlink fronthaul coordination. An extension to generic non-orthogonal fronthaul channels is worth pursuing.

\bibliography{arXiv}
\bibliographystyle{ieeetr}

\end{document}